\documentclass[acmtog]{acmart}

\acmSubmissionID{132}

\setcopyright{acmcopyright}
\acmJournal{TOG}
\acmYear{2022}
\acmVolume{41}
\acmNumber{6}
\acmArticle{197}
\acmMonth{12} 
\acmDOI{10.1145/3550454.3555433}

\usepackage{booktabs} %
\usepackage{color}
\usepackage{xcolor}
\usepackage{graphicx}
\usepackage{booktabs}
\usepackage{amsmath, amsthm, amsfonts, amssymb}
\usepackage{mathrsfs}
\usepackage{textcomp}
\usepackage{epstopdf}
\usepackage{multirow}
\usepackage{wrapfig}
\usepackage{subfig}
\usepackage{bbm}

\usepackage{algorithm}  
\usepackage{algorithmicx}  
\usepackage{algpseudocode}

\usepackage{ragged2e}
\usepackage[normalem]{ulem}

\definecolor{green}{rgb}{0, 0.5, 0}
\definecolor{orange}{rgb}{0.6, 0.3, 0.1}
\definecolor{red}{rgb}{1.0, 0.0, 0.0}
\definecolor{teal}{rgb}{0.0, 0.4, 0.4}
\definecolor{purple}{rgb}{0.65,0,0.65}
\definecolor{saffron}{rgb}{0.8,0.55,0.1}
\definecolor{turquoise}{rgb}{0.0,0.5,0.5}
\definecolor{brown}{rgb}{0.5, 0.16, 0.16}
\definecolor{brickred}{rgb}{.6, .2 .1}
\definecolor{coral}{rgb}{1,0.45,0.33}
\definecolor{newcolor}{rgb}{.8,.349,.1}

\newcommand{\rev}[1]{{\color{black}#1}}

\newcommand{\hui}[1]{{\color{black}#1}}

\newcommand{\rz}[1]{{\color{black}#1}}

\newcommand{\ylr}[1]{{\color{black}#1}}

\newcommand{\ie}{i.e.}
\newcommand{\eg}{e.g.}

\begin{document}

\title{Learning Reconstructability for Drone Aerial Path Planning}

\author{Yilin Liu}
\email{whatsevenlyl@gmail.com}
\affiliation{%
	\institution{Shenzhen University}
	\country{China}	
}	
\author{Liqiang Lin}
\email{liniquie@gmail.com}
\affiliation{%
	\institution{Shenzhen University}
	\country{China}	
}
\author{Yue Hu}
\email{hytraveler2000@gmail.com}
\affiliation{%
	\institution{Shenzhen University}
	\country{China}	
}
\author{Ke Xie}
\email{ke.xie.siat@gmail.com}
\affiliation{%
	\institution{Shenzhen University}
	\country{China}	
}
\author{Chi-Wing Fu}
\email{philip.chiwing.fu@gmail.com}
\affiliation{%
	\institution{The Chinese University of Hong Kong}
	\country{China}	
}
\author{Hao Zhang}
\email{hao.r.zhang@gmail.com}
\affiliation{%
	\institution{Simon Fraser University}
	\country{Canada}	
}
\author{Hui Huang}
\email{hhzhiyan@gmail.com}
\authornote{Corresponding author: Hui Huang (hhzhiyan@gmail.com)}
\affiliation{%
	\department{College of Computer Science \& Software Engineering}
	\institution{Shenzhen University}
	\country{China}	
}

\renewcommand\shortauthors{Y. Liu, L. Lin, Y. Hu, K. Xie, C. Fu, H. Zhang, and H. Huang}

\begin{abstract}
We introduce the first {\em learning-based reconstructability predictor\/} to improve view and path planning for
large-scale 3D urban scene acquisition using unmanned drones. In contrast to previous heuristic approaches, our method learns a model that {\em explicitly\/} predicts how well a 3D urban scene will be reconstructed from a set of viewpoints. To make such a model trainable and simultaneously applicable to drone path planning, we simulate the proxy-based 3D scene reconstruction during training to set up the prediction. Specifically, the neural network we design is trained to predict the scene reconstructability as a function of the {\em proxy geometry\/}, a set of viewpoints, and optionally a series of scene images acquired in flight. To reconstruct a new urban scene, we first build the 3D scene proxy, then rely on the predicted reconstruction quality and uncertainty measures by our network, based off of the proxy geometry, to guide the drone path planning. We demonstrate that our data-driven reconstructability predictions are more closely correlated to the true reconstruction quality than prior heuristic measures. Further, our learned predictor can be easily integrated into existing path planners to yield improvements. Finally, we devise a new iterative view planning framework, based on the learned reconstructability, and show superior performance of the new planner when reconstructing both synthetic and real scenes.
\end{abstract}



\begin{teaserfigure}
	\centering
	\includegraphics[width=\linewidth]{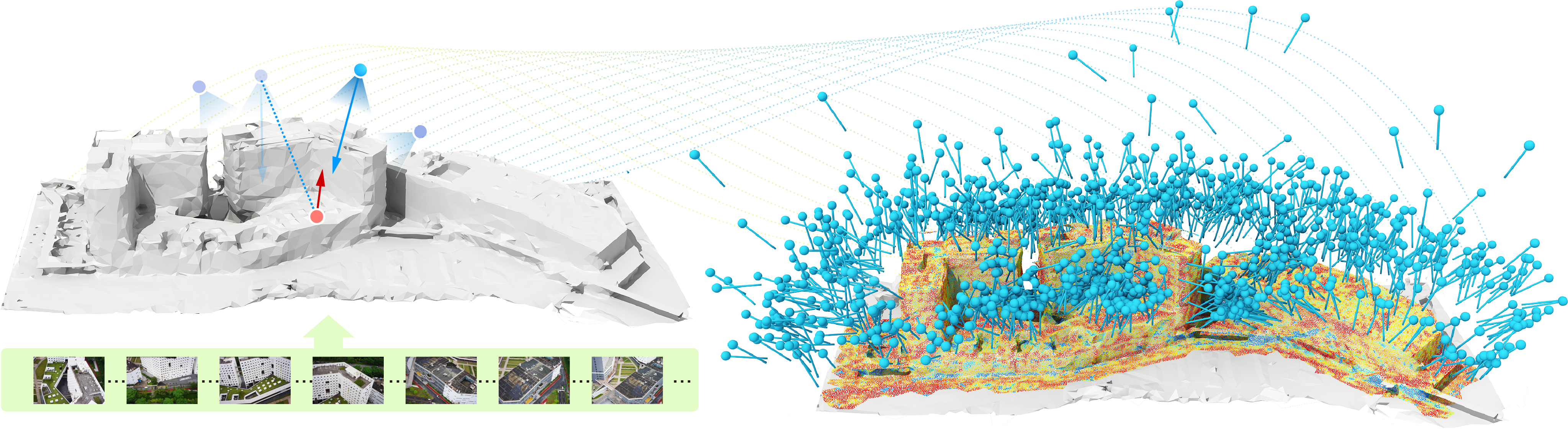}
	\vspace*{-5.5mm}
	\caption{We train a neural network to predict reconstructability for drone path planning during 3D urban scene acquisition. The prediction is based on a rough scene proxy, a set of viewpoints (bluish dots on the top), and optionally a series of images captured during the pre-flight pass, as shown on the left. The network learns both image features and viewpoint features from the perspective of sample points (red dot) on the proxy, while
	the predicted reconstructability (color) map, shown on the right, guides our iterative view planner to execute the onsite drone view acquisition for 3D reconstruction.}
	\label{fig:teaser}
	\vspace*{1mm}
\end{teaserfigure}

\maketitle

\section{Introduction}
\label{sec:intro}

Aerial path planning for large-scale 3D urban scene acquisition using unmanned drones has gained much attention
recently~\cite{Roberts17ICCV,Hepp18Plan3D,Wolfgang18SIGA,DroneFly21,DronePath21,DroneScan20}. The ultimate 
goal of the planning problem is to best reconstruct the 3D scenes in terms of completeness and accuracy while respecting physical 
constraints imposed by the drones' flight speed %
and battery life. 
As the scene complexity scales up, it becomes more difficult to model the spatial relations between a scene and its viewpoints, and more importantly, the {\em uncertainty\/} surrounding scene reconstruction quality which forms the foremost criterion for path planning. Since the ground-truth 3D scene is unknown during planning, 
the reconstruction quality, or specifically, scene {\em reconstructability\/} with respect to a set of viewpoints of the drones, must be estimated using imperfect input data.

Current approaches to estimating reconstructability all rely on heuristics in one form or another. 
Some account for scene coverage~\cite{Giang2021SequentialDC,Schmid20ActivePlanning} or viewpoint correlations~\cite{FurukawaCSS10,KochRS19} without scene reconstruction. Others optimize path planning based on a coarse {\em proxy reconstruction\/}~\cite{Roberts17ICCV,Wolfgang18SIGA,Hepp18Plan3D,DronePath21} 
obtained by an initial drone fly through.
However, scene coverage and view correlation represent measures that are only {\em relevant to\/} reconstruction quality; they do not explicitly model it or strictly validate heuristic designs against it, even when the data is complete. On the other hand, the pre-constructed proxy geometries are typically coarse and inaccurate and these inaccuracies are easily propagated to the coverage and correlation estimates so as to misguide the ensuing path planning.

We formulate reconstructability estimation as a {\em predictive\/} task and introduce the first
data-driven approach to {\em learn\/} reconstructability for drone aerial path planning. 
\rev{More formally, reconstructability measures how well the local area around a sample point in the input 3D scene can be 
reconstructed from information captured at a set of viewpoints.
In contrast to previous heuristic approaches, our learned model {\em explicitly\/} predicts
how well a 3D urban scene will be reconstructed.} To make such a model trainable and simultaneously applicable 
to drone path planning, we simulate the {\em proxy-based\/} 3D scene reconstruction during training to set up the
prediction. Specifically, \rev{our training data contains ground-truth reconstructability values per sample point on proxy geometry subject 
to a given viewpoint set, so that we can train the network to predict final ``reconstruction \ylr{quality}'' at the sample points, as
a function of the {\em proxy geometry\/} and the viewpoints.} 
Note that technically, the reconstruction \ylr{quality} is measured against the ground-truth scene, but it is unknown during inference;
only the scene proxy is available.

\rev{As an extension to our learning framework, we leverage additional image inputs to refine the
reconstructability prediction, since high-quality images can often be acquired during the drone pre-flight. 
These images can better capture scene details than would be possible via proxy construction, 
thus helping to} predict \textit{uncertainty-aware} inaccuracies over the proxy geometry.
To reconstruct a new scene, we first build the 3D proxy, then rely on the network predictions
and uncertainty measures, based off of the proxy geometry, to guide the \rev{view planner
to find a set of viewpoints to \ylr{maximize} the predicted reconstruction \ylr{quality} for 3D scene reconstruction.}

Our learning model consists of an attention-based view fusion network for reconstructability prediction. As such, the
influence of the different factors related to a single viewpoint on scene reconstruction, such as viewing distances and viewing angles 
(with respect to surface normals over the scene geometry), together with the correlations between multiple viewpoints, such as their scale differences and  
baselines (i.e., distances), are all adaptively adjusted by the learned parameters. Further, we develop another attention-based {\em image\/}
fusion network to implicitly model the uncertainty of the scene geometry with respect to the acquired image observations
and refine the {\em spatial\/} reconstructability that is learned from spatial relations between the viewpoints and scene proxy; see e.g., Fig.~\ref{fig:teaser}.

We demonstrate the effectiveness of our learning framework through extensive experiments, both quantitatively and qualitatively. 
We verify that our learned reconstructability more closely correlates to the true reconstruction quality than prior 
heuristic measures. Important for immediate practical impact, our reconstructability predictor can be easily integrated into state-of-the-art 
path planners~\cite{Wolfgang18SIGA,DroneScan20}, leading to improved quality for large-scale 3D urban scene acquisition.

Finally, we complete the loop by devising a new {\em iterative viewpoint optimization\/} framework,
based on the learned reconstructability, to further improve path planning. Specifically, we adjust the current
viewpoints along with the path planning process to attain better reconstructability, where the adjustments
include viewpoint insertion near under-reconstructed regions, deletion of redundant views, and 
altering the position and orientation of existing views. We show that our new adaptive scheme, built on a
more accurate reconstructability prediction, can help escape local minima during path planning, a reoccurring
issue which has challenged existing planners. The new planner exhibits superior reconstruction performance
over existing methods on both synthetic and real scenes.

\section{Related Work} 
\label{sec:related}

Unmanned drones have been widely employed for urban scene acquisition due to their maneuverability and large fields of view. During an acquisition, a drone usually flies along a pre-computed path, which is generated to optimize a quality measurement. Predominantly, such a measurement is related to the completeness and accuracy of
the 3D urban scene to be reconstructed, %
i.e., reconstructability, with respect to a set of viewpoints and paths selected. However, since the ground-truth geometry is unknown, all methods must estimate the reconstructability measure. 

\subsection{Estimates of reconstructability}

\paragraph{View coverage and uncertainty.}
One line of approaches to path optimization is based on view coverage~\cite{Giang2021SequentialDC, Schmid20ActivePlanning} by a depth sensor. Schmid et al.~\shortcite{Schmid20ActivePlanning} proposed a spatial uncertainty measure based on viewing distance. A rapid random search tree was developed to maintain the measure, facilitating the search for an optimal scanning path. Song et al.~\shortcite{SongKJ20} divide the acquisition process into two steps: global planning and local inspection. During global planning, they also rely on a scene uncertainty measure to generate a rough initial path. This is followed by solving a set cover problem to optimize the local viewpoints so as to capture more geometric details.
Note that when planning paths to better cover the 
target urban %
scenes, these methods all consider view coverage with respect to individual viewpoints. On the other hand, methods based on multi-view stereo (MVS),~\eg,~\cite{Wolfgang18SIGA, FurukawaCSS10,PengICRA19}, often need to account for spatial relations between the viewpoints, since the errors stemming from triangulation and feature matching all depend on the relative positions of the viewpoints and the scene geometry.

\paragraph{View correlation.}
To this end, several measures have been proposed to characterize correlations between a {\em pair\/} of viewpoints in order to model reconstruction quality. %
Furukawa et al.~\shortcite{FurukawaCSS10} assumed that the quality measure follows a piecewise Gaussian distribution, which 
depends on viewpoint baselines and the pixel densities.
Smith et al.~\shortcite{Wolfgang18SIGA} decomposed this measure into two components, which respectively account for feature matching and triangulation in the context of MVS. Furthermore, their work defines reconstructability as an accumulative product of Gaussian functions, which are defined in terms of viewpoint baselines, view distances, and viewing angles. In addition, Peng and Isler~\shortcite{PengICRA19} also considered the impact of different view sampling rates when performing feature matching and dense reconstruction. Finally, Koch et al.~\shortcite{KochRS19} factored in viewpoint overlaps during optimization.

While the above methods all consider view pairs when measuring reconstruction errors, in real MVS reconstruction, multiple viewpoints visible to a surface point would contribute to its nearby reconstruction. %
Roberts et al.~\shortcite{Roberts17ICCV} extended the correlation model to encompass a {\em set\/} of viewpoints. Specifically, they proposed a measure based on spherical integration, which considers the impact of all visible viewpoints. The integral function is related to viewing distances, viewpoint baselines, and viewing angles. However, like other methods, which also consider these unary and relational viewpoint attributes, various assumptions have to be made, resulting in a variety of parameters that are difficult to tune in practice.

\paragraph{Scene proxy.}
During path planning, most methods up to now obtain the various measures with respect to a rough scene proxy obtained either via a rapid pre-fly and rough reconstruction~\cite{Roberts17ICCV,Wolfgang18SIGA,Hepp18Plan3D,DronePath21}, i.e., the scene proxy, or by an extraction from geological features~\cite{DroneScan20}. The inaccuracies or uncertainties of the proxy scene geometry, especially pertaining to surface normal estimation, would significantly impact the view planning. Peng and Isler~\shortcite{PengICRA19} developed a three-step scene reconstruction method, by iteratively finding reconstructed regions that have the lowest confidence and conducting path planning for them, to improve reconstruction quality. However, this method needs several drone flights to obtain a satisfactory reconstruction, leading to high acquisition costs.

\paragraph{Data-driven methods.}
Recently, the rapid proliferation of 3D scene datasets~\cite{chang2015shapenet,huang2019apolloscape,Knapitsch2017,VGFNet21,UrbanScene3D} 
have enabled data-driven methods to model correlations between viewpoints and scene geometry. 
Genova et al.~\shortcite{Genova2017LearningWT} proposed such a method for view set selection, whereby a set of views are generated for a synthetic dataset to
match the content statistics of a set of example images.
Sun et al.~\shortcite{SunCVPR21} designed a neural network to model the visibility and quality of viewpoints, turning the traditional discrete viewpoint optimization problem into a continuous one.
At last, Hepp et al.~\shortcite{Hepp18ECCV} utilized voxel maps for encoding the correlation between viewpoints and scene geometry, allowing one to predict the quality of the next viewpoint.
However, these methods only model visibility or the quality of a single view, and they are still limited to depth-based reconstruction.

\rev{
On the other hand, transformer~\cite{vaswani_attention_2017} is an effective means for extracting the correlation between data.
It has been extensively employed in machine translation~\cite{vaswani_attention_2017}, as well as stereo depth estimation~\cite{li_revisiting_2021} and multi-view reconstruction~\cite{ding_transmvsnet_2022}.
In our work, we adopt transformers to fuse the geometric relations between multiple viewpoints and the image information in the reconstructability measurement.
}

\subsection{Path planning}

Based on the various reconstructability estimates, different path planners have been proposed to optimize viewpoint configuration for urban scene acquisition, where
most of them~\cite{Roberts17ICCV,Wolfgang18SIGA,DroneScan20} uniformly initialize a set of viewpoints as optimization candidates.
Roberts et al.~\shortcite{Roberts17ICCV} first coarsen the view optimization by determining the optimal direction for each viewpoint and finding the additive approximation of it. 
Then a standard integer linear program solver is employed to solve such an {\em orienteering\/} problem to obtain the optimal trajectory. 
Similarly, Smith et al.~\shortcite{Wolfgang18SIGA} generate an initial trajectory at a fixed height with nadir view orientation. 
Then they use their reconstructability measurement as the objective function to iteratively identify whether a new position and orientation for each viewpoint are better. 
To this end, they resort to the Nelder-Mead method to find the global minimum of their objective function. 
Hepp et al.~\shortcite{Hepp18Plan3D} voxelized the 3D safe space and define their objective function as the information gain towards an unknown environment.
More recently, Zhou et al.~\shortcite{DroneScan20} leverage a dense initialization of the viewpoints and assume the initialization to be perfect but redundant for the reconstruction. 
Then they define the view contribution of each viewpoint according to their reconstructability measure and iteratively reduce redundant viewpoints to obtain the optimal subset of viewpoints. 

On the other hand, some path planners directly generate a trajectory without any viewpoint initialization. 
Zhang et al.~\shortcite{DronePath21} maintain and expand a rapidly-exploring random tree of the scene to directly obtain a more efficient trajectory with sufficient reconstructability. 
Liu et al.~\shortcite{DroneFly21} generate and update the image acquisition path in real time through certain pre-defined trajectory patterns on a coarse scene proxy. 
However, the binary visibility function from a viewpoint to a surface point and the correlation between different viewpoints make this problem non-convex, hence hard to optimize in practice. Minimizing the objective function in an iterative way is prone to be stuck in local minima.

In our work, we develop the first {\em learned\/} reconstructability predictor, and along with an associated view optimization scheme, we can improve the performance of existing path planners.

\section{Overview}
\label{sec:overview}

\begin{figure*}[t!]
	\centering
	\includegraphics[width=1\textwidth]{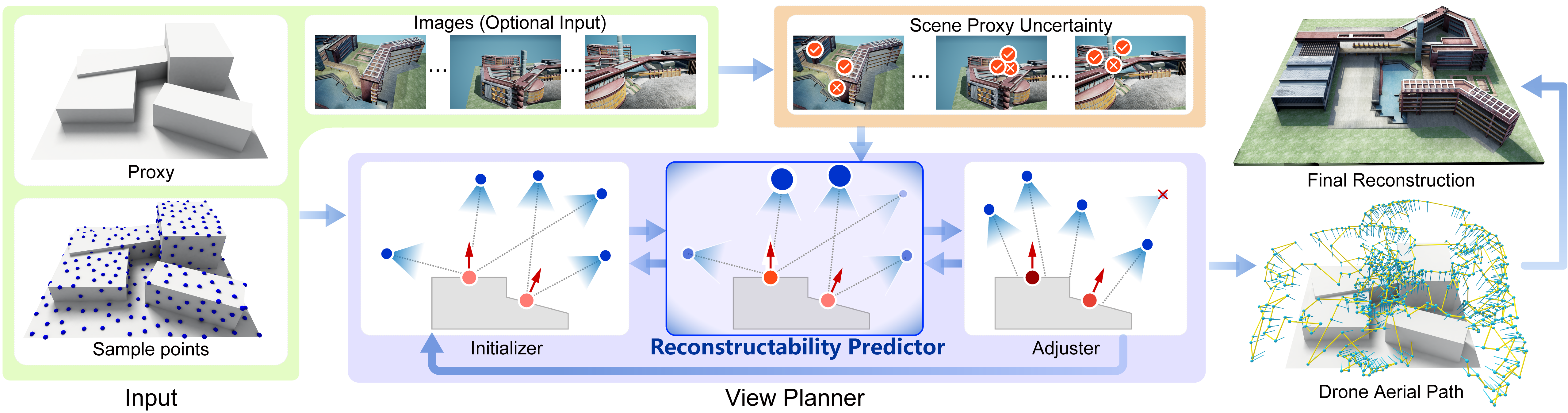}
	\caption{
	\textit{Left:} our approach takes the proxy geometry of the target scene, sample points (in red)
	on the proxy, images captured by a pre-flight, and the camera poses as input.
	\textit{Middle:} our approach consists of a \emph{reconstructability predictor} and a \emph{view planner}. The \emph{reconstructability predictor} first extracts spatial features between the sample points and the viewpoints (in blue). If the images and camera poses in the pre-flight are available, our \emph{reconstructability predictor} can extract the scene proxy uncertainty near this sample point from the images and predict the uncertainty-aware spatial reconstructability for the view planner. The \emph{view planner} takes the output of the \emph{reconstructability predictor} and iteratively optimizes the number and poses of each viewpoint to maximize their ability to reconstruct the target scene. \textit{Right:} the trajectory obtained by our method and the corresponding reconstruction result.
	}
	\label{fig:overview}
\end{figure*}

Our learning-based framework consists of two phases: (i) {\em training phase\/} and (ii) {\em inference phase\/}.
In the training phase, we prepare training data using the UrbanScene3D dataset ~\cite{UrbanScene3D} and train our neural network model to predict scene reconstructability on the proxy geometry against ground-truth information extracted from the data. Then, in the inference phase, we can integrate our trained network model, as a \textit{reconstructability predictor}, into existing view planners for calculating the scene reconstructability.  Further, to address the limitations of the existing planners, we formulate a new view planning framework that combines the strengths of the existing ones
to produce view configurations.

\vspace*{-3pt}
\paragraph{Network inputs}
To train our neural network, we first prepare network inputs based on a given set of viewpoints in the free space and a given set of sample points on the scene proxy; see Fig.~\ref{fig:overview} (left).
On the one hand, we explore every visible (sample) point-view pair, considering their locations and orientations, and encode their spatial relation geometrically as a 5D point-view feature vector.
These spatial relations provide hints on how well the image captured at each
viewpoint would contribute to reconstructing the local geometry surrounding the sample point.
On the other hand, we extract {\em image features\/} from each pre-captured image upon its availability.

\vspace*{-3pt}
\paragraph{Network predictions}
From the network inputs, we design our neural network to first extract point-view spatial features through MLPs (Multi Layer Perceptions); then, our network adopts a transformer encoder to explore the correlations across sample points and viewpoints, enabling us to better fuse features from different sample points for \rz{predicting} the spatial reconstructability at each sample point on the scene proxy (Sec.~\ref{sec:method_geometric}).
On the other hand, upon the availability of pre-captured images, our network also extracts image features and fuses these features with the point-view features to enable us to predict uncertainty-aware spatial reconstructability (Sec.~\ref{sec:method_uncertainty}).
By this means, we can better account for 
the \emph{inaccuracy} in
scene proxy in the reconstructability prediction.

\vspace*{-3pt}
\paragraph{View planning}
Last, we integrate our trained network into existing view planners as a measure for scene reconstructability (Sec.~\ref{sec:integration}).
By doing so, we found limitations of two state-of-the-art planners,~\cite{Wolfgang18SIGA} and~\cite{DroneScan20}, on optimizing viewpoints for scene acquisition.
Hence, we further formulate a new view planner (Sec.~\ref{sec:new_planner}), collectively combining their complementary strengths by iteratively initializing, eliminating, and adjusting viewpoints, as guided by our trained network, to obtain a view configuration with maximized reconstructability; see Fig.~\ref{fig:overview}.

\vspace*{3pt}
In the end, we \hui{evaluate our reconstructability predictor and view planner in both unseen synthetic and real scenes.
Results presented in Sec.~\ref{sec:results} show that the proposed reconstructability predictor can better reflect the final reconstruction quality, while the view planner can produce drone acquisition trajectories that lead to better reconstruction results compared to the previous methods~\cite{Wolfgang18SIGA,DroneScan20,DronePath21,DroneFly21}}.

\section{Reconstructability}
\label{sec:recon}

\begin{figure*}[t]
	\centering
	\includegraphics[width=1\textwidth]{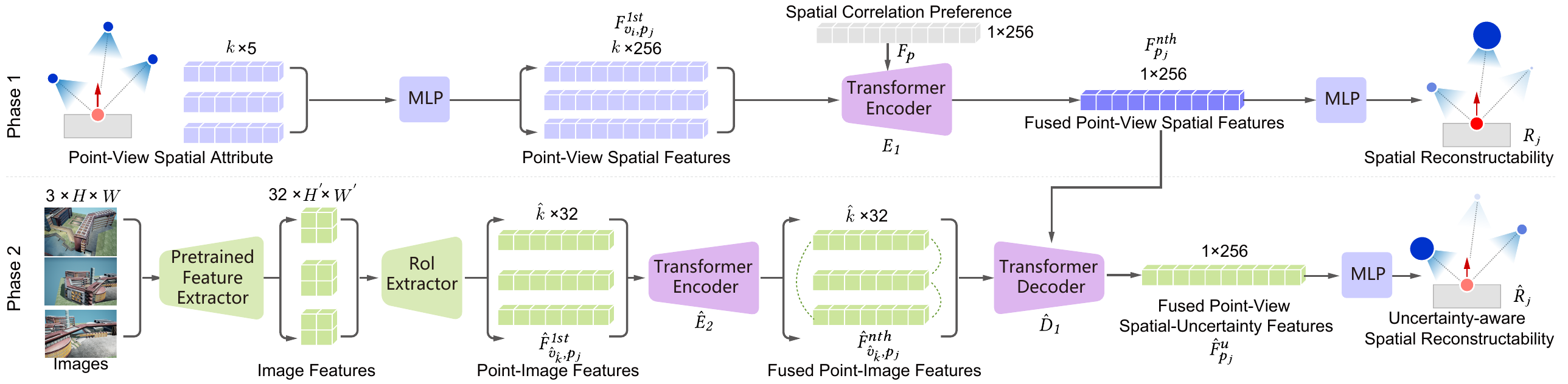}
	\caption{The input to our network is a set of point-view spatial attributes for each sample point, and optionally images from a pre-flight. Based on the geometric characteristics between the sample points and viewpoints, our network automatically extracts the contribution of each viewpoint and predicts the \textit{spatial reconstructability} of each sample point. If images from the pre-flight are available, our network can further extract the uncertainty of the proxy geometry near each sample point and predict the \textit{uncertainty-aware reconstructability} to enhance the subsequent view planning process.
	}
	\label{fig:network}
\end{figure*}

\rev{
Reconstructability is an essential measure of how well a set of viewpoints reconstructs the target scene.
Both our proposed and the existing planners~\cite{Wolfgang18SIGA,DroneScan20,DronePath21} rely on it to optimize the view planning.
Yet, unlike existing works, 
we define the reconstructability measure by formulating a learning approach and considering the ultimate goal of reconstructability, which is to enhance the quality of the final scene reconstruction.
So, we adopt the reconstruction error metrics, accuracy and completeness, from Smith et al.~\shortcite{Wolfgang18SIGA}, and define the reconstructability term in our framework to be inversely proportional to the reconstruction error (Sec.~\ref{sec:method_training}).
Hence, when we train our framework, minimizing the training loss would then drive our framework to learn to predict high (low) reconstructability for scene regions with low (high) reconstruction error.
As a result, we can employ our framework to predict learned reconstructability in view planners to better estimate the final reconstruction quality.
}

The key to 
\rev{formulating the learning approach}
is to find the relation between the viewpoints and the scene geometry.
Given $N$ viewpoints $\{ v_i \}_{i=1}^N$ and sample point $p_j$ on proxy geometry, 
we want to learn function $G_s: (\mathbb{R}^{6 \times N}, \mathbb{R}^{6}) \rightarrow \mathbb{R}^{1}$ that predicts
\begin{align}
& R_j = G_{s}( \{ v_i \} , p_j ),
\end{align}
where each view $v_i$ consists of a position and an orientation;
each sample point $p_j$ consists of a position and a normal vector; and
$R_j$ is the spatial reconstructability of point $p_j$ that measures how well $p_j$ 
can be reconstructed by the viewpoints $\{ v_i \}$.

Also, we want to consider the uncertainty of the given scene geometry when predicting the reconstructability, \emph{if} some images of the target scene are given. Compared to the spatial relations function $G_s$ we modeled above, images provide rich texture information, which can help improve both the \textit{reconstructability} prediction and the subsequent path planning. Specifically, \rev{given $L$ existing RGB images \{$\hat{I_l}\}_{l=1}^L$ and their poses \{$\hat{v}_l\}_{l=1}^L$,} we want to learn another function $\hat{G}_{s}: (\mathbb{R}^{L \times 3 \times H \times W}, \mathbb{R}^{6 \times L}, \mathbb{R}^{6 \times N}, \mathbb{R}^{6}) \rightarrow \mathbb{R}^{1}$ that predicts 
\begin{align}
    & 
    \hat{R}_{j} = \hat{G}_{s}( \{ \hat{I}_l \} , \{ \hat{v}_l \} , \{ v_i \} , p_j ),
\end{align}
where $H$ and $W$ are the size of images and 
$\hat{R}_{j}$ 
denotes the refined \textit{reconstructability}, considering the potentially inaccurate geometry nearby. We show our network structure in Fig.~\ref{fig:network}. In the following section, we 
present
the geometric representation of $R_s$ in Sec.~\ref{sec:method_geometric} and how to compute the refined measurement 
$\hat{R}_{s}$ in Sec.~\ref{sec:method_uncertainty}. 
\rev{Also, we give the training details and a more specific calculation for our reconstructability measure in Sec.~\ref{sec:method_training}.}

\subsection{Spatial Reconstructability}
\label{sec:method_geometric}

When measuring the quality of a set of viewpoints, we want to formulate $G_{s}$ to extract the relation among the viewpoints, sample points, and \textit{reconstructability}. Unlike the previous method~\cite{Wolfgang18SIGA}, which assumes that $G_{s}$ follows 
a Gaussian distribution associated with
some manually-defined parameters, we use a data-driven method to learn this function.

\paragraph*{Individual view-point feature extraction}
For each sample point $p_j$, 
\setlength{\intextsep}{0pt}
\setlength{\columnsep}{3pt}
\begin{wrapfigure}[7]{r}{2cm}
    \centering
    \includegraphics[width=\linewidth]{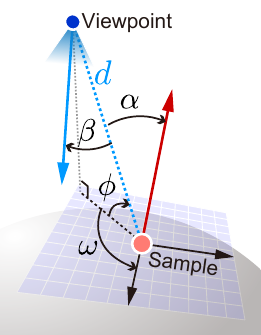}
    \centering
    \label{wrapfig:view_attribute}
\end{wrapfigure}
we first locate all visible viewpoints at $p_j$ and calculate their spatial 
attributes with respect to $p_j$. As shown in the inset figure below,
we consider the following five elements as the spatial features between point $p_j$ and view $v_i$ to predict the spatial reconstructability: the 
local spherical coordinates $(\omega,\phi,d)$ of the viewpoint with respect to the sample point, angle $\alpha$ between the sample point normal and the direction from the sample point to the viewpoint, and angle $\beta$ between the viewpoint's viewing direction and the direction from the viewpoint to sample point. \rev{In practice, angles $\alpha$ and $\phi$ are complementary, so we only need to calculate one of them.}

To better encode the individual influence of viewpoint $v_i$ on sample point $p_j$, \rev{we use an MLP (Multi Layer Perception) module to map the 5D spatial features to a 256D latent vector $F_{v_i,p_j}^{\mathit{1st}}, which$
represents the influence of \textit{individual} viewpoint on the associated local scene geometry during the reconstruction.}

\paragraph*{Viewpoint feature fusion}
In the multi-view stereo (MVS) pipeline, viewpoints are highly coupled. The change in relative position and orientation of the viewpoints has a big impact on the final reconstruction~\cite{FurukawaCSS10,Wolfgang18SIGA}. So, we further extract a higher-order correlation between the viewpoints. Unlike the previous method~\cite{Wolfgang18SIGA}, which exhaustively computes the feature of every view pair to approximate this correlation, we adopt a transformer encoder to learn the correlations among viewpoints and sample points, extract the contribution of each viewpoint, and then transform the individual feature $F_{v_i,p_j}^{\mathit{1st}}$ of \rev{$K$ visible viewpoints} to a fused point-view spatial feature $F_{p_j}^{\mathit{nth}}$. Specifically, we train the transformer encoder, \rev{$E_1: \mathbb{R}^{(K+1) \times 256} \rightarrow \mathbb{R}^{1 \times 256}$, with
\begin{align}
    F_{p_j}^{\mathit{nth}} = E_1( F_p, \{ F^{1st}_{v_i,p_j} \}_{i=1}^K )
\end{align}
where the query, key, and value are all from $K$ individual feature\rev{s} 
$\{ F_{v_i,p_j}^{\mathit{1st}} \}_{i=1}^K$. 
}
Similar to DETR~\cite{detr}, we also use trainable parameter $F_p: \mathbb{R} ^ {1 \times 256}$ to represent the spatial correlation preference. \rev{We stack $F_p$ on the input features to form a $K+1$ tensor at the beginning and extract the fused point-view spatial feature $F_{p_j}^{\mathit{nth}}$ from $F_p$ after the fusion.}

\paragraph*{Spatial reconstructability}
Last, we use a standard MLP to learn to determine \rev{
the spatial reconstructability $R_j$ from the fused point-view spatial feature $F_{p_j}^{\mathit{nth}}$.}

\subsection{Uncertainty-aware Spatial Reconstructability}
\label{sec:method_uncertainty}

The quality of the final reconstructed model is not only related to the geometric relation between the viewpoints and the scene geometry but also influenced by the surface appearance.
More importantly, the scene geometry we used in the above computation is usually reconstructed by a quick pre-flight pass~\cite{Wolfgang18SIGA,Hepp18ECCV,DroneScan20}, which provides only coarse or even inaccurate geometry information.
Hence, we train another function 
$\hat{G_s}$, 
which 
leverages the images captured from the pre-flight to further refine the reconstructability $R_j$ into 
the \textit{uncertainty-aware spatial reconstructability} $\hat{R}_{j}$.

\paragraph*{Individual point-image feature extraction}
\rev{
    For each pre-captured image, we first extract their 32D latent code using a pre-trained convolutional neural network~\cite{cascade_mvs}. In order to obtain the image feature of each sample point $p_j$ on the scene geometry, we project point $p_j$ on the image and use a feature interpolation operator~\cite{maskrcnn} to extract the individual feature $\hat{F}_{\hat{v}_{\hat{k}},p_j}^{1st}: \mathbb{R} ^ {32 \times 1}$ of sample point $p_j$ in $\hat{k}th$ viewpoint $\hat{v}_{\hat{k}}$. 
}

\paragraph*{Point-image feature fusion}
As the sample point can be visible at multiple viewpoints, we adopt another transformer encoder, $\hat{E}_2: \mathbb{R} ^ {\hat{K} \times 32} \rightarrow \mathbb{R} ^ {\hat{K} \times 32}$, to 
correlate and fuse the
$\hat{K}$ individual point-image feature $\hat{F}_{\hat{v}_{\hat{k}},p_j}^{1st}$ to produce the fused image feature $\hat{F}_{\hat{v}_{\hat{k}},p_j}^{nth}$ over all the visible viewpoints $\hat{V}_k$.

\paragraph*{Uncertainty-aware spatial reconstructability fusion}
Then, we can use the fused point-image feature of point $p_j$ to refine the spatial reconstructability $R_j$ that we predicted before. We adopt transformer decoder, $\hat{D}_1: (\mathbb{R} ^ {\hat{K} \times 32}, \mathbb{R} ^ {1 \times 256}  ) \rightarrow \mathbb{R} ^ {1 \times 256}$, 
to extract the importance of the fused point-image feature $\hat{F}_{\hat{v}_{\hat{k}},p_j}^{nth}$ to the spatial feature $F_{p_j}^{\mathit{nth}}$ and output the fused feature $\hat{F}_{p_j}^{u}: \mathbb{R} ^ {1 \times 256}$ of point $p_j$ with
\begin{align}
    \hat{F}_{p_{j}}^{u} = \hat{D}_1 ( F_{p_{j}}^{\mathit{nth}}, \hat{F}_{\hat{v}_{\hat{k}},p_{j}}^{nth})
\end{align}
where we use spatial feature $F_{p_j}^{\mathit{nth}}$ we predicted before as the query tensor to represent the pure spatial feature around this point and
the fused image feature 
$\hat{F}_{\hat{v}_{\hat{k}},p_j}^{nth}$
as the key and value to refine the feature, injecting semantics around point $p_j$ into the prediction.

\paragraph*{Uncertainty-aware spatial reconstructability}
Finally, we use an MLP
\rev{module
to predict the final uncertainty-aware spatial reconstructability $\hat{R}_j$ for point $p_j$ from the fused feature $\hat{F}_{p_j}^{u}$.}

\begin{figure}[t]
    \centering
    \includegraphics[width=1\linewidth]{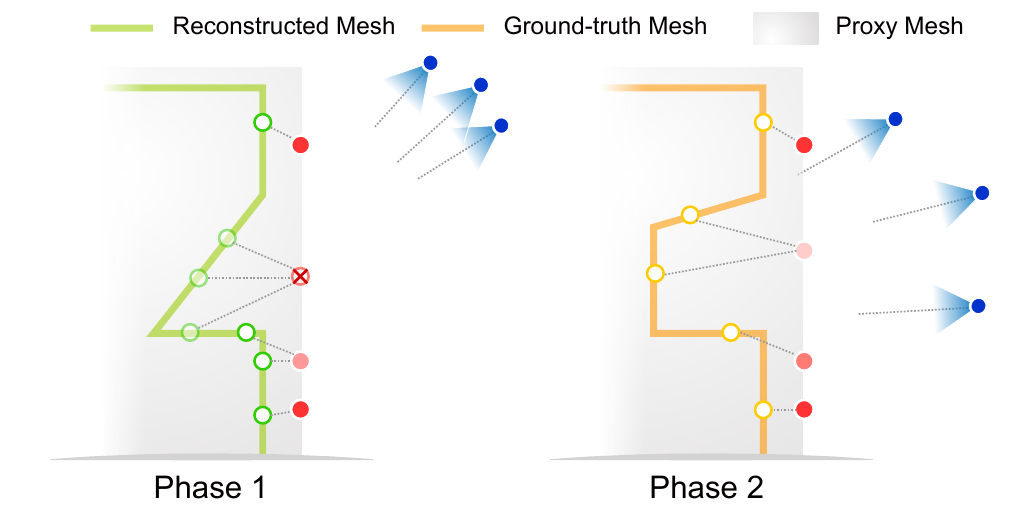}
    \caption{We use different projection mechanisms for generating ground-truth labels (targets) in different training phases. 
    Blue dots are viewpoints and red dots are sample points on proxy, where deeper red means the sample point has higher reconstructability.
    In \textit{Phase 1}, we use the reconstruction accuracy between the reconstructed model 
    and proxy 
    to generate the ground-truth 
    \textit{reconstructability} for training our network, enabling it to learn the spatial correlation between the sample points and viewpoints.
    In \textit{Phase 2}, we use the reconstruction completeness between the ground-truth surface and proxy surface to generate the target for our learned reconstructability, encouraging the network model to encode the uncertainty measurement in the reconstructability prediction.
    }
    \label{fig:data_generation}
\end{figure}

\begin{figure*}[t!]
	\centering
	\includegraphics[width=1\linewidth]{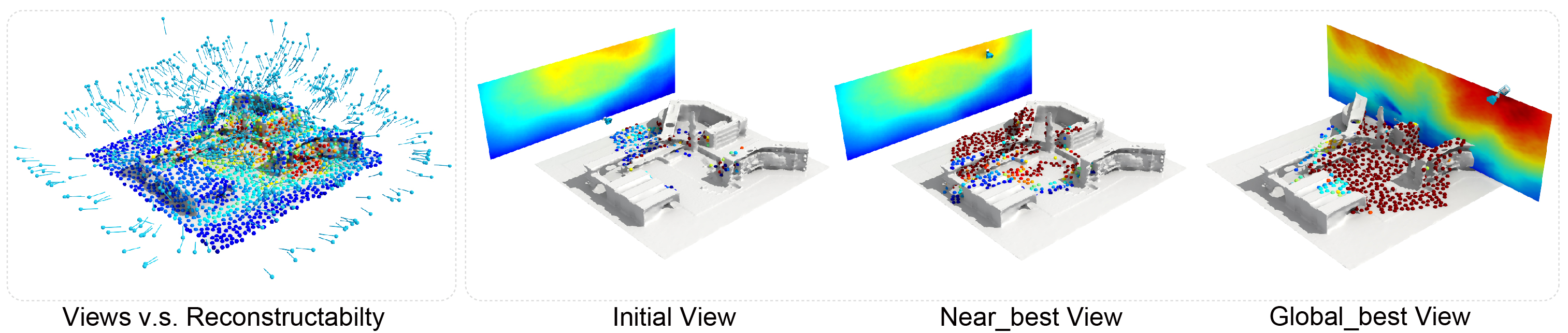}
	\caption{Here we show a common phenomenon: \textit{local minima} may easily occur in the optimization process of previous view planners.  The left figure shows 
	all
	viewpoints and sample points with their associated
	reconstructability~\cite{Wolfgang18SIGA} in the current configuration.
	As shown in the second figure, we select a specific viewpoint as our analysis target.
	We change its position in the plane and calculate the reconstructability \emph{increment} at that position.
    For each position, we randomly sample 128 directions and find the best one with the highest reconstructability to the sample points. 
    \rev{To better visualize the calculated value nearby, we only show a section of the reconstructability field.}
    The ``best'' place, or the local minimum, near this viewpoint can be found in the third figure, where the reconstructability of all sample points can be maximized. However, the right figure shows that the actual best place, or the global minimum, appears far away from the initial view location. In practice, it is very difficult to 
    find the local minimum during the planning, since previous planners~\cite{Wolfgang18SIGA,DroneScan20} only leverage local information to optimize viewpoints.
    \rev{Our proposed view planning framework samples more viewpoints near poorly-reconstructed regions and reduces the number of viewpoints near well-reconstructed regions to help the planner escape from the local minima.}
    }
	\label{fig:local_minimum}
\end{figure*}

\subsection{\rev{Training}}
\label{sec:method_training}

\rev{
While the reconstructability measures in existing works are well-defined and easy to obtain, taking them as target to train our framework will only drive our framework to predict reconstructability that mimics the existing measures.
Our learning approach goes beyond existing works by considering the ultimate goal of reconstructability,~\ie, to enhance the final reconstruction quality.
Hence, we explicitly supervise the training of our network by simulating proxy-based acquisitions for various scenes and also the reconstruction process with planned viewpoints, such that during the training, we can define the target value of the network-predicted reconstructability based on the ground-truth reconstruction errors.

Specifically, we prepare our training data using UrbanScene3D~\cite{UrbanScene3D}, which consists of different scenes and different levels of the proxy~\cite{DroneScan20,Wolfgang18SIGA}, as well as trajectories and associated reconstruction results from different path planners~\cite{DroneScan20,Wolfgang18SIGA,DronePath21}.
As \rev{described} in Sec.~\ref{sec:overview}, our network input is a set of 5D relative information between the sample point and viewpoints, the visible pre-captured images, and their poses.
We can easily obtain such input data from UrbanScene3D.
However, training our framework is still hard, since it involves multiple data sources and configurations.
Also, the image data is optional, as it may not be available during path planning.
So, we split our training process into two phases and adopt different strategies to prepare training data in each phase.
In particular, we use the \textit{fine} and \textit{inter} levels of proxy to train our framework in phase 1, since they have smaller difference from the ground-truth model.
Then in phase 2, we use all four proxy levels to train the uncertainty-aware reconstructability predictor.

\paragraph*{Phase 1 training}
The goal of phase 1 is to model the pure geometric function $G_s$, which associates only with the relative information geometrically between the sample point on proxy and the viewpoints.
From UrbanScene3D, we can obtain the \textit{reconstruction accuracy}~\cite{Wolfgang18SIGA}, which is defined on the sample points on the reconstructed model and measured by the shortest distance to the ground-truth surface.
Then, we can project the reconstruction accuracy to the sample points on the proxy geometry, such that we can estimate the reconstruction accuracy on the proxy.
In detail, for each sample point $p_j$ on proxy, we find a set of nearest sample points $\{p_q\}$ on the reconstructed mesh within distance threshold $\tau$. 
Then, by averaging accuracy $acc_q$ of each point $p_q$, we can implicitly 
encode 
the potential error %
near each sample point $p_j$ on proxy:
\begin{align}
R_j^g = \frac { |\{p_q\}| } { \sum_{q \in \{p_q\}} {acc_q}  },
\end{align}
where $R_j^g$ is used as the target of our reconstructability measure when training our framework.
Also, we discard sample points on proxy
with distances to the reconstructed model larger than $\tau$.
This helps avoid taking proxy inaccuracy into our framework, as
these sample points are likely located on inaccurate parts of the proxy.

\paragraph*{Phase 2 training}
Then, phase 2 further considers the pre-captured images~\cite{Wolfgang18SIGA}, enabling us to 
predict uncertainty-aware reconstructability.
Specifically, we define the training target directly using the \textit{reconstruction completeness}~\cite{Wolfgang18SIGA},
which is defined on the ground-truth model and measured by the shortest distance to the reconstructed model.
Note that the proxy may differ significantly from the ground-truth model, due to insufficient data when preparing the proxy; see Fig.~\ref{fig:data_generation}.
Hence, phase 2 further encodes such uncertainty into the reconstructability prediction.

\paragraph*{Training details}
As Fig.~\ref{fig:network} shows, we set all hidden dimensions in our network to 256, except for the dimension of the image features (\ie,~32), which is from the pre-trained model~\cite{cascade_mvs}.
The input images are resized to $800 \times 600$ 
\rev{with color values normalized to $[0,1]$.}
We use the standard RoI-Align operator~\cite{maskrcnn} to extract the feature of each specific point in the given image. The MLP module we adopted consists of a linear layer and an ReLU activation. For data generation, we use $\tau=20cm$ in all our experiments. Also, we use an \textit{L1} loss to train the network.
The whole training process takes about 10 hours on an RTX 3090Ti GPU.
}

\section{View Planner}
\label{sec:planner}

In this section, we first discuss the integration of our reconstructability predictor with existing view planners in Sec.~\ref{sec:integration}.
Due to the non-linearity of the optimization process, existing planners may easily be stuck at local minima.
So, in Sec.~\ref{sec:new_planner}, we further present a new view planning framework to overcome this problem.

\begin{figure*}[t!]
	\centering
	\includegraphics[width=1\linewidth]{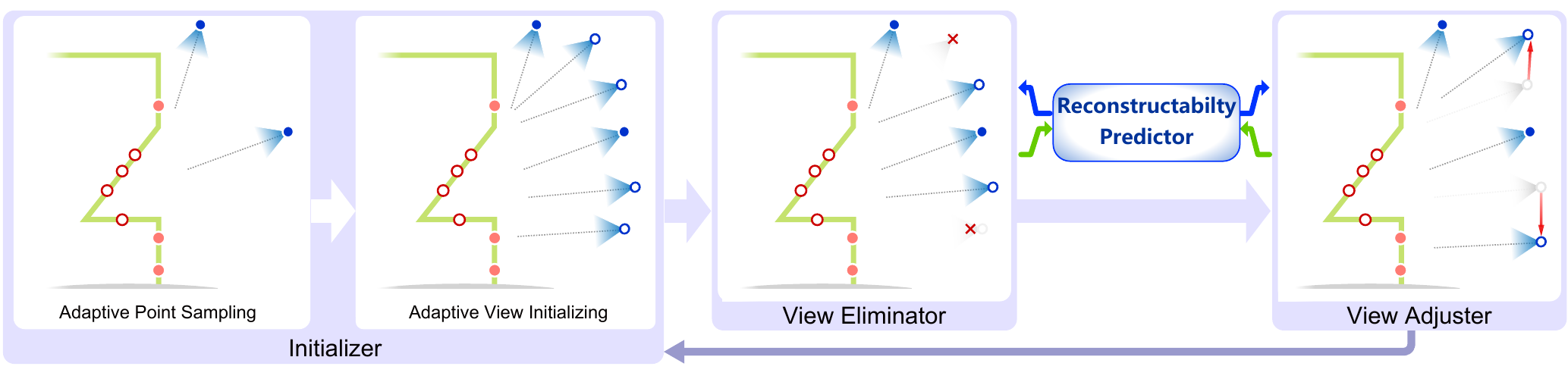}
	\caption{Our proposed view-planning framework. Different from the existing ones, which are based on either
	view elimination~\cite{DroneScan20} or view adjustment~\cite{Wolfgang18SIGA}, our \textit{initializer} iteratively finds sample points with 
	low reconstructability and try to allocate more views around these regions.
	The whole planner runs in an iterative way, helping it to escape from local minima during the optimization.}
	\label{fig:planner}
\end{figure*}

\subsection{Integration with Existing Planners}
\label{sec:integration}

Our reconstructability measure 
can be easily integrated into existing planners~\cite{Wolfgang18SIGA,DroneScan20}.
Smith et al.~\shortcite{Wolfgang18SIGA} 
can use our reconstructability to find an optimal viewpoint set that maximizes the reconstructability of sample points. We can readily 
replace the reconstructability calculation using our method.
More specifically, we can use our predicted reconstructability to identify if a new viewpoint configuration is better than the original one when executing the downhill simplex method in their method. So, we can use the same view planner to minimize the same objective.

Also, we can integrate our method into the view planner in~\cite{DroneScan20}.
We use our predicted reconstructability to calculate the view redundancy of a given view configuration and perform the subsequent min-max view reduction. In each iteration, we follow~\cite{DroneScan20} to 
select the viewpoint with the highest redundancy and remove it temporally. Then, we can use our reconstructability predictor to test if any sample point receives a reconstructability lower than the threshold. %
If the test passes all 
associated sample points, we can remove the viewpoint permanently.

During the view planning, the objective function usually involves the relative position between the sample points and viewpoints, as well as between different 
viewpoints. Thus, the objective function is 
highly
non-linear. Since the existing planners mostly optimize the view configuration in an iterative manner, they may easily 
fall into local minima.
We further show this phenomenon in Fig.~\ref{fig:local_minimum}.

Previous planners 
often fail to escape from local minima. Smith et al.~\shortcite{Wolfgang18SIGA} only find view candidates near the current position, preventing it from moving redundant views near poorly-reconstructed regions.
As for Zhou et al.~\shortcite{DroneScan20}, local minimum occurs when the view initialization is not perfect. However, it is hard to obtain a perfect view initialization for scenes with complex structures and occlusions. Also, having more initialized viewpoints will increase the computational burden on the visibility test and reconstructability calculation, as the complexity of reconstructability calculation is 
$O(|V|^2)$, where $|V|$ is the number of viewpoints.

Yet, we find that these two methods can complement each other. The local view adjustment from Smith et al.~\shortcite{Wolfgang18SIGA} can help increase the reconstructability even with a poor view initialization from Zhou et al.~\shortcite{DroneScan20} by finding a better position and orientation near each viewpoint.
On the other hand, the view initialization and elimination from Zhou et al.~\shortcite{DroneScan20} can help the local view adjustment from Smith et al.~\shortcite{Wolfgang18SIGA} escape from local minima. Particularly, it helps avoid redundant views 
near well-reconstructed regions and allocate more views around regions with poor reconstructability.

\subsection{New Planner}
\label{sec:new_planner}

To this end, we develop a new view planning framework, which iteratively initializes, eliminates, and adjusts viewpoints to obtain a view configuration with maximum reconstructability. Our view planner consists of an initializer, a view eliminator, and a view adjuster, as shown in Fig.~\ref{fig:planner}. Compared with the previous planners, our 
planner optimizes viewpoints in an iterative manner and can better escape from local minima during the optimization through adaptive point sampling and view initializing.

\paragraph*{Adaptive points sampling}
Zhou et al.~\shortcite{DroneScan20}
first collect a dense set of viewpoints as the initialization 
and assume a perfect initialization.
So, the planner only needs to reduce the number of viewpoints. However, 
it is hard to obtain a perfect viewpoint initialization,
as 
the proxy geometry is usually coarse and inaccurate.

\begin{figure}[t]
    \centering
    \includegraphics[width=1\linewidth]{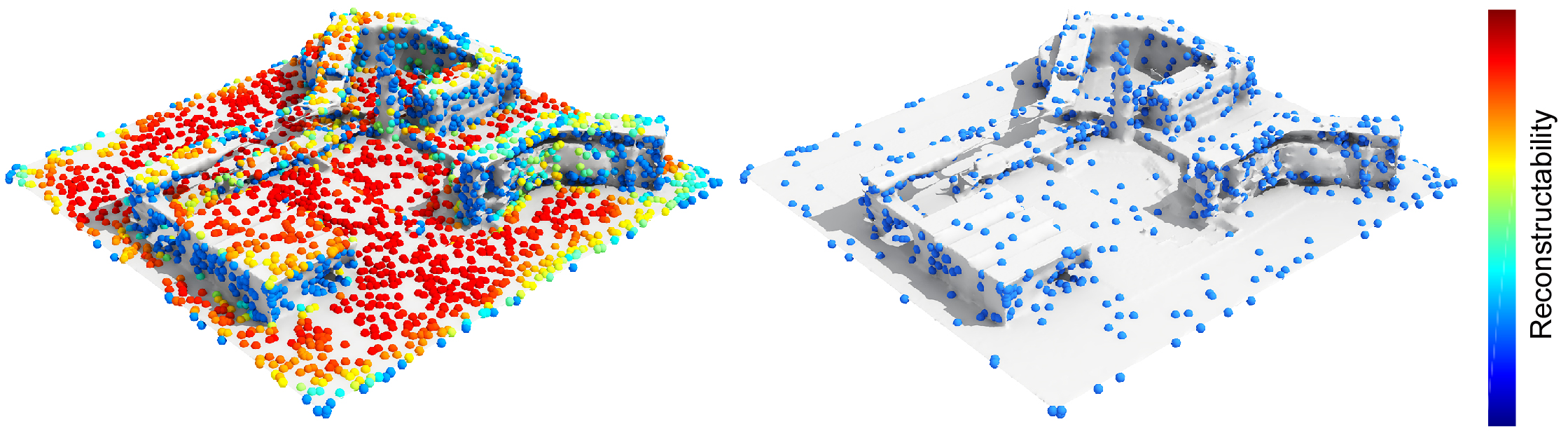}
    \caption{Our sampling mechanism that generates the optimized target at each iteration. The left figure shows the current reconstructability of each sample point in the scene. The probability of each point to be sampled in the current iteration is calculated according to its reconstructability. The right figure shows the sampling result. These points are used to initialize, eliminate, and adjust viewpoints in the current iteration. This strategy helps the planner focus on regions that are hard to be reconstructed.}
    \label{fig:planner_sampling}
\end{figure}

Specifically, we select sample points in an adaptive manner.
The probability $Prob_{p_j}$ of selecting sample point $p_j$ is calculated based on its current reconstructability $R_j$ (or $\hat{R}_{j}$, if images are provided):
\begin{align}
    Prob_{p_j} = \frac{\sum_{q \in P_n} {\rev{\frac{1}{R_q}}
    e^{-d_q}
    }}{|P_n|},
\end{align}
where $P_n$ is the set of sample points nearest to $p_j$ and $d_q$ is the distance from nearest sample point $q$ to point $p_j$.
By this distance-weighted average, we can find regions that are not well-reconstructed and sample more points in them; see Fig.~\ref{fig:planner_sampling} 
for an illustration.

Compared with previous methods, which use uniform sample points as the optimization target, our proposed adaptive sampling can enable us to obtain better resolutions for regions that are previously hard to be reconstructed.

In each iteration, we sample $N$ points on the proxy surface as the optimization target according to the above probability. Note that the following view initialization elimination and adjustment will only be performed on the selected sample points.

\begin{figure}[t]
	\centering
	\includegraphics[width=1\linewidth]{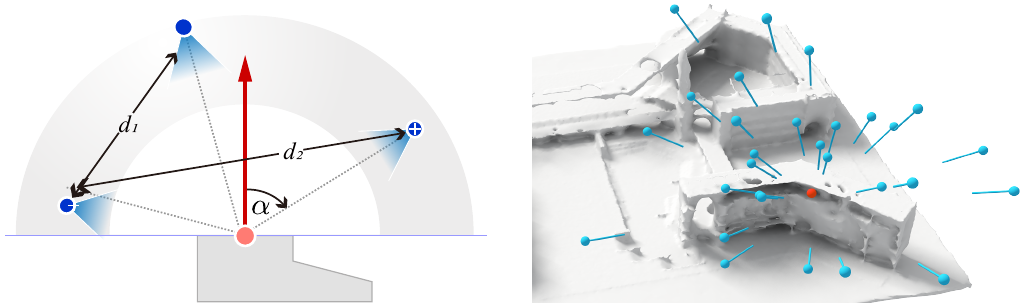}
	\caption{We initialize viewpoints according to the normal vector of each sample point and the existing visible viewpoints toward this point. Specifically, we want the initialized viewpoint closer to the normal direction of the sample point (smaller $\alpha$), while being far away from the existing viewpoints (larger $d_1,d_2$). The left shows an example of our view initialization.
	}
	\label{fig:planner_init}
\end{figure}

\begin{figure*}[t!]
	\centering
	\includegraphics[width=1\linewidth]{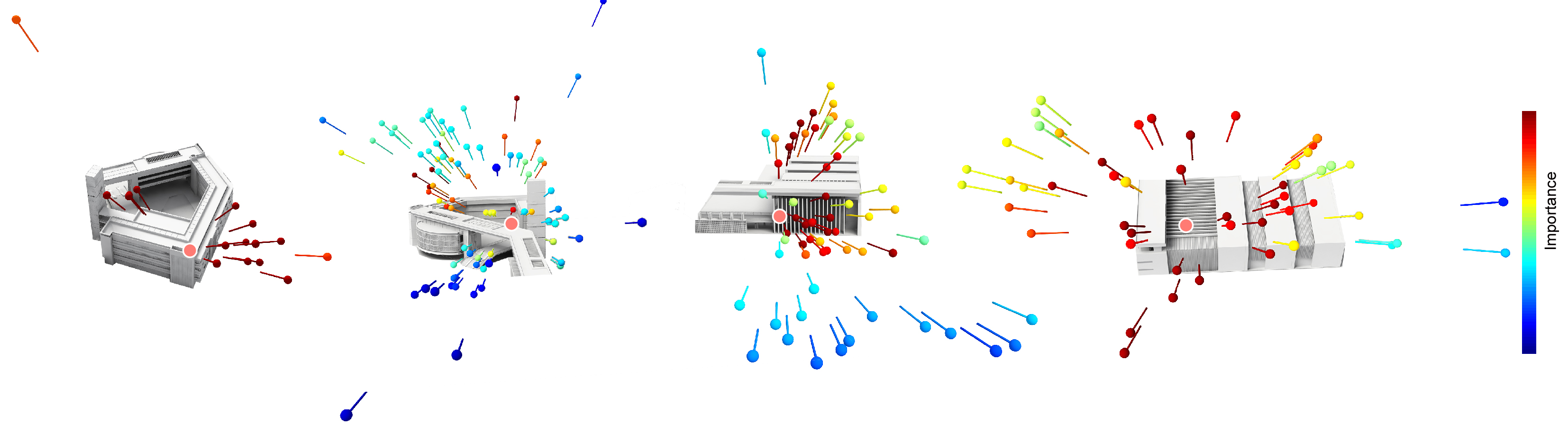}
	\caption{The contribution of each viewpoint when predicting the reconstructability. Viewpoints with high contribution are marked red. Note that we only use the final reconstructability as the supervised signal during training. The contribution automatically extracted from the network indicates that viewpoints from far away have less impact on results, and viewpoints with appropriate baselines have higher weights when computing reconstructability.}
	\label{fig:eva_contribution}
\end{figure*}

\begin{figure}[t]
	\centering
	\includegraphics[width=1\linewidth]{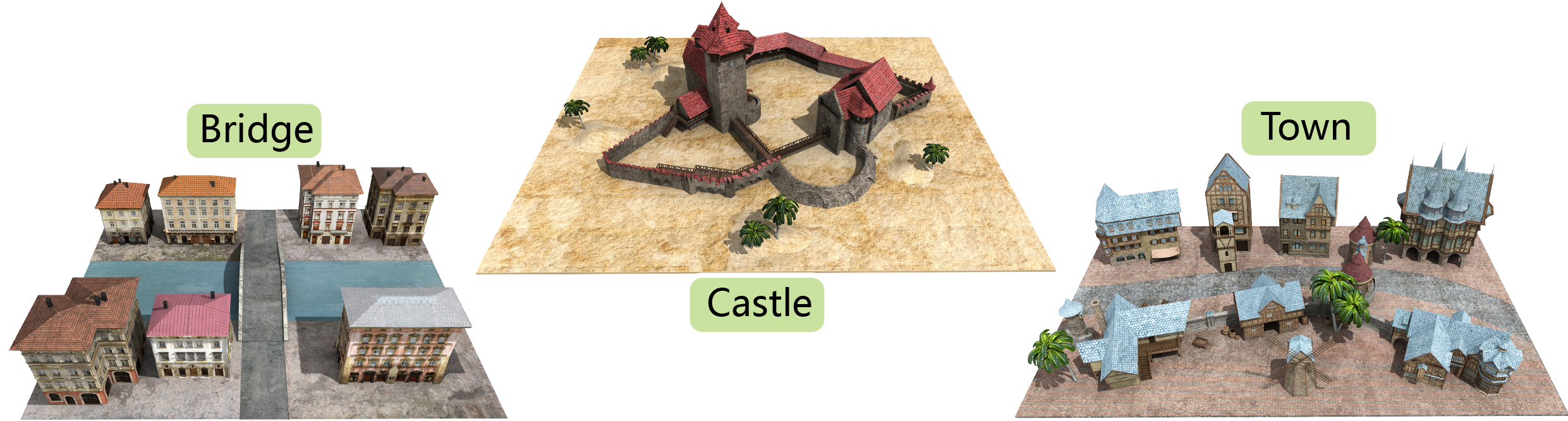}
	\caption{The training set for our reconstructability predictor. We use \textit{Bridge, Castle, Town} from UrbanScene3D dataset~\cite{UrbanScene3D}, which contains 72 trajectories and the corresponding reconstruction results. For testing, we use a completely new scene: \textit{School} to evaluate our predictor. Also, we use  another dataset~\cite{Wolfgang18SIGA} and three real scenes to test the integration of our reconstructability predictor with various path planners~\cite{Wolfgang18SIGA,DroneScan20}.}
	\label{fig:eva_training_set}
\end{figure}

\paragraph*{Adaptive view initialization}

For each sample point on the proxy surface, we create
a set of viewpoints 
as
a local initialization around the sample point.
Zhou et al.~\shortcite{DroneScan20} directly extend the normal vector of each sample point to a specific view distance and place a viewpoint towards the sample point.  However, complex geometric structures and occlusions on the proxy geometry will simply break this initialization, as shown in Fig.~\ref{fig:planner_init}. Also, the viewing direction of the initialized viewpoint in Zhou et al.~\shortcite{DroneScan20} 
always points to the associated sample point. Such a setting could be optimal for observing the associated sample point, but there should be better choices by considering multiple adjacent sample points. Instead, we randomly sample $M_m$ viewpoints in a hemisphere around each sample point. 
\rev{We then filter the visible viewpoints according to the proxy and}
add the best $M_b$ viewpoints to our viewpoint set. The weight of each sample viewpoint $v_m$ is calculated by
\begin{align}
    Score_{v_m} = dot(v_m - x_j, n_j) * \min_{v_v \in V_v} dot (v_m - x_j, v_v - x_j),
\end{align}
where $x_j$, $n_j$ are the position and normal vector of sample point $p_j$; and
$V_v$ is the existing set of visible viewpoints at point $p_j$.
The calculation 
encourages the viewpoint to be closer to the normal vector of point $p_j$, while further away from the existing viewpoints.

\paragraph*{View elimination and adjustment}

Based on the initialized viewpoints, we use Zhou et al.~\shortcite{DroneScan20} to compute the redundancy of each viewpoint and remove the redundant ones. Similar to Smith et al.~\shortcite{Wolfgang18SIGA}, we also adjust the viewpoints to further increase the reconstructability of the sample points after the view elimination.

\section{Results and Evaluation}
\label{sec:results}

We start with an analysis of the proposed reconstructability predictor in Sec.~\ref{sec:eva_recon}, by reporting the correlation factor between the reconstructability predicted by different methods and the final reconstruction quality. Next, we integrate our reconstructability predictor into several existing view planners to
demonstrate improved final reconstruction quality in Sec.~\ref{sec:eva_integration}. This is followed by experimenting with and evaluating the new view planner we propose, in Sec.~\ref{sec:eva_planner}. Finally, Sec.~\ref{sec:eva_real} presents results from our full view planning and reconstruction pipeline on three real scenes.

\begin{table*}[t!]
    \newcommand{\tabincell}[2]{\begin{tabular}{@{}#1@{}}#2\end{tabular}}
    \newcommand{\mcl}[1]{\multicolumn{1}{|c|}{#1}}
    \newcommand{\tg}[1]{\multirow{12}{*}{\rotatebox{90}{#1}}}
    \newcommand{\B}[1]{\textbf{#1}}
    \caption{Quantitative comparison between different reconstructability estimates on the test scene: \textit{School}, \textit{without} image inputs. Higher Spearman correlation factor indicates better prediction.  \textit{Visible number} denotes the correlation factor between the reconstruction quality and the number of visible viewpoints at each sample point. Compared with the two baselines, the reconstructability predicted by our method better matches the final reconstruction quality.  }
    \centering
    \label{table:spearman_p0}
    \begin{tabular}{ccccccccc}
        \toprule
        \multirow{2}{*}{Planner}      & \multirow{2}{*}{Overlap} & \multirow{2}{*}{Proxy} & \multirow{2}{*}{ Image (\#)} & Ours        & \multicolumn{2}{c}{Smith et al.} & \multicolumn{2}{c}{Visible Number}                          \\ \cmidrule(r){5-5} \cmidrule(r){6-7} \cmidrule(r){8-9}
                                      &                          &                        &                        & Spearman    & Spearman                         & Inc.                               & Spearman & Inc.        \\ \hline \hline
        \multirow{2}{*}{Smith et al.} & \multirow{2}{*}{70}      & Inter                  & 559                    & \B{32.42\%} & 17.83\%                          & 81.83\%                            & 22.83\%  & 42.01\%     \\
                                      &                          & Fine                   & 559                    & \B{43.35\%} & 36.66\%                          & 18.25\%                            & 36.15\%  & 19.92\%     \\\hline
        \multirow{2}{*}{Smith et al.} & \multirow{2}{*}{90}      & Inter                  & 559                    & \B{14.79\%} & 10.31\%                          & 43.45\%                            & 12.32\%  & 20.05\%     \\
                                      &                          & Fine                   & 559                    & \B{34.94\%} & 30.86\%                          & 13.22\%                            & 26.68\%  & 30.96\%     \\\hline
        \multirow{2}{*}{Zhou et al.}  & \multirow{2}{*}{70}      & Inter                  & 342                    & \B{25.49\%} & 17.62\%                          & 44.67\%                            & 14.37\%  & 77.38\%     \\
                                      &                          & Fine                   & 518                    & \B{44.51\%} & 41.48\%                          & 7.30\%                             & 37.84\%  & 17.63\%     \\\hline
        \multirow{2}{*}{Zhou et al.}  & \multirow{2}{*}{90}      & Inter                  & 595                    & \B{23.02\%} & 11.48\%                          & 100.52\%                           & 12.02\%  & 91.51\%     \\
                                      &                          & Fine                   & 1243                   & \B{38.79\%} & 32.07\%                          & 20.95\%                            & 29.82\%  & 30.08\%     \\ \hline
        \multirow{2}{*}{Zhang et al.} & \multirow{2}{*}{70}      & Inter                  & 330                    & \B{30.53\%} & 27.83\%                          & 9.70\%                             & 27.23\%  & 12.12\%     \\
                                      &                          & Fine                   & 518                    & \B{36.35\%} & 30.58\%                          & 18.87\%                            & 34.37\%  & 5.76\%      \\\hline
        \multirow{2}{*}{Zhang et al.} & \multirow{2}{*}{90}      & Inter                  & 570                    & \B{32.88\%} & 24.63\%                          & 33.50\%                            & 31.46\%  & 4.51\%      \\
                                      &                          & Fine                   & 1043                   & \B{49.53\%} & 44.57\%                          & 11.13\%                            & 43.39\%  & 14.15\%     \\ \hline
        Average Inc.                  &                          &                        &                        &             &                                  & \B{33.62\%}                        &          & \B{30.51\%} \\
        \bottomrule
    \end{tabular}
\end{table*}

\begin{table*}[t!]
    \newcommand{\tabincell}[2]{\begin{tabular}{@{}#1@{}}#2\end{tabular}}
    \newcommand{\mcl}[1]{\multicolumn{1}{|c|}{#1}}
    \newcommand{\tg}[1]{\multirow{24}{*}{\rotatebox{90}{#1}}}
    \newcommand{\B}[1]{\textbf{#1}}
    \caption{Quantitative comparison between different reconstructability estimates on the \textit{School} scene, \textit{with} image inputs, where the image features were computed for images captured during drone pre-flight. Again, our predictor performs better than the baselines even with inaccurate proxies (\textit{Coarse, Inter}). }
    \centering
    \label{table:spearman_p1}
    \begin{tabular}{ccccccccc}
        \toprule
        \multirow{2}{*}{Planner}      & \multirow{2}{*}{Overlap} & \multirow{2}{*}{Proxy} & \multirow{2}{*}{ Image (\#)} & Ours        & \multicolumn{2}{c}{Smith et al.} & \multicolumn{2}{c}{Visible Number}                             \\ \cmidrule(r){5-5} \cmidrule(r){6-7} \cmidrule(r){8-9}
                                      &                          &                        &                        & Spearman    & Spearman                         & Inc.                               & Spearman    & Inc.        \\ \hline \hline
        \multirow{4}{*}{Smith et al.} & \multirow{4}{*}{70}      & Box                    & 559                    & \B{26.89\%} & 7.30\%                           & 268.36\%                           & 7.69\%      & 249.67\%    \\
                                      &                          & Coarse                 & 559                    & \B{27.07\%} & 9.16\%                           & 195.52\%                           & 8.89\%      & 204.50\%    \\
                                      &                          & Inter                  & 559                    & \B{36.02\%} & 16.88\%                          & 113.39\%                           & 23.50\%     & 53.28\%     \\
                                      &                          & Fine                   & 559                    & \B{55.51\%} & 53.12\%                          & 4.50\%                             & 52.19\%     & 6.36\%      \\\hline
        \multirow{4}{*}{Smith et al.} & \multirow{4}{*}{90}      & Box                    & 559                    & \B{15.69\%} & 8.66\%                           & 81.18\%                            & 11.55\%     & 35.84\%     \\
                                      &                          & Coarse                 & 559                    & \B{7.65\% } & 1.13\%                           & 576.99\%                           & 1.08\%      & 608.33\%    \\
                                      &                          & Inter                  & 559                    & \B{29.85\%} & 11.62\%                          & 156.88\%                           & 20.27\%     & 47.26\%     \\
                                      &                          & Fine                   & 559                    & \B{45.12\%} & 40.06\%                          & 12.63\%                            & 37.34\%     & 20.84\%     \\\hline
        \multirow{4}{*}{Zhou et al.}  & \multirow{4}{*}{70}      & Box                    & 416                    & 23.23\%     & 21.54\%                          & 7.85\%                             & \B{24.00}\%  & -3.21\%     \\
                                      &                          & Coarse                 & 330                    & \B{25.50\%} & 17.78\%                          & 43.42\%                            & 19.99\%     & 27.56\%     \\
                                      &                          & Inter                  & 342                    & \B{41.30\%} & 19.38\%                          & 113.11\%                           & 18.32\%     & 125.44\%    \\
                                      &                          & Fine                   & 518                    & \B{56.35\%} & 55.23\%                          & 2.03\%                             & 49.40\%     & 14.07\%     \\\hline
        \multirow{4}{*}{Zhou et al.}  & \multirow{4}{*}{90}      & Box                    & 614                    & \B{16.58\%} & 11.23\%                          & 47.64\%                            & 11.05\%     & 50.05\%     \\
                                      &                          & Coarse                 & 570                    & \B{22.25\%} & -2.02\%                          & 1201.49\%                          & 4.89\%      & 355.01\%    \\
                                      &                          & Inter                  & 595                    & \B{37.12\%} & 6.16\%                           & 502.60\%                           & 12.40\%     & 199.35\%    \\
                                      &                          & Fine                   & 1243                   & \B{47.01\%} & 40.41\%                          & 16.33\%                            & 39.90\%     & 17.82\%     \\\hline
        \multirow{4}{*}{Zhang et al.} & \multirow{4}{*}{70}      & Box                    & 518                    & \B{29.79\%} & 16.00\%                          & 86.19\%                            & 14.30\%     & 108.32\%    \\
                                      &                          & Coarse                 & 330                    & \B{23.57\%} & 14.32\%                          & 64.59\%                            & 14.68\%     & 60.56\%     \\
                                      &                          & Inter                  & 330                    & \B{44.12\%} & 33.76\%                          & 30.69\%                            & 35.25\%     & 25.16\%     \\
                                      &                          & Fine                   & 518                    & 57.04\%     & 52.36\%                          & 8.94\%                             & \B{57.88\%} & -1.45\%     \\\hline
        \multirow{4}{*}{Zhang et al.} & \multirow{4}{*}{90}      & Box                    & 614                    & \B{23.78\%} & 14.60\%                          & 62.88\%                            & 11.10\%     & 114.23\%    \\
                                      &                          & Coarse                 & 570                    & 9.92\%      & 6.90\%                           & 43.77\%                            & \B{12.13\%} & -18.22\%    \\
                                      &                          & Inter                  & 570                    & \B{45.42\%} & 33.03\%                          & 37.51\%                            & 37.82\%     & 20.10\%     \\
                                      &                          & Fine                   & 1043                   & \B{60.82\%} & 59.55\%                          & 2.13\%                             & 60.45\%     & 0.61\%      \\\hline
        Average Inc.                  &                          &                        &                        &             &                                  & \B{153.36\%}                       &             & \B{96.73\%} \\
        \bottomrule
    \end{tabular}
\end{table*}

\begin{table*}[t!]
    \newcommand{\tabincell}[2]{\begin{tabular}{@{}#1@{}}#2\end{tabular}}
    \newcommand{\mcl}[1]{\multicolumn{1}{|c|}{#1}}
    \newcommand{\tg}[1]{\multirow{6}{*}{\rotatebox{90}{#1}}}
    \newcommand{\B}[1]{\textbf{#1}}
    \caption{\rev{Quantitative evaluation, on reconstruction quality using F-score, Precision, and Recall, of integrating our reconstructability predictor into two different view planners~\cite{DroneScan20,Wolfgang18SIGA}, either using image inputs or not. The method being compared to, marked by ``Smith", employed reconstructability estimates by the method in Smith et al.~\shortcite{Wolfgang18SIGA} to guide the view planning for 3D scene reconstruction.}}
    \centering
    \label{table:comparison_recon_fpr}
    \begin{tabular}{cccccccccc}
        \toprule
        \multirow{2}{*}{Proxy} & \multirow{2}{*}{Recon.} & \multicolumn{4}{c}{Smith et al. Planner} & \multicolumn{4}{c}{Zhou et al. Planner}                                                                                                                    \\ \cmidrule(r){3-6} \cmidrule(r){7-10}
                               &                         & Image (\#)                               & F-score$\uparrow$                       & Precision$\uparrow$ & Recall$\uparrow$ & Image (\#) & F-score$\uparrow$ & Precision$\uparrow$ & Recall$\uparrow$ \\ \hline\hline
        \multirow{3}{*}{Box}   & Smith                   & 1,770                                     & 31.8491                                 & 44.8348             & 24.6963          & 806        & 25.6819           & 42.4220             & 18.4151          \\
                               & Ours (w/o)              & 1,717                                     & 32.5645                                 & 45.6230             & 25.3178          & 699        & 27.6641           & 44.2545             & \B{20.1210}      \\
                               & Ours                    & 1,610                                     & \B{34.5483}                             & \B{52.4389}         & \B{25.7598}      & 796        & \B{28.1596}       & \B{48.5530}         & 19.8303          \\ \hline
        \multirow{3}{*}{Inter} & Smith                   & 714                                      & 34.3846                                 & 50.9298             & \B{25.9533}      & 825        & 29.2754           & 46.7762             & 21.3045          \\
                               & Ours (w/o)              & 747                                      & \B{34.6040}                             & \B{52.3987}         & 25.8315          & 747        & 29.9230           & 47.4468             & 21.8522          \\
                               & Ours                    & 696                                      & 34.4410                                 & 52.1732             & 25.7047          & 696        & \B{32.4972}       & \B{53.8454}         & \B{23.2709}      \\
        \bottomrule
    \end{tabular}
\end{table*}

\begin{table*}[t!]
    \newcommand{\tabincell}[2]{\begin{tabular}{@{}#1@{}}#2\end{tabular}}
    \newcommand{\mcl}[1]{\multicolumn{1}{|c|}{#1}}
    \newcommand{\tg}[1]{\multirow{6}{*}{\rotatebox{90}{#1}}}
    \newcommand{\B}[1]{\textbf{#1}}
    \caption{\rev{Quantitative comparison between the different planners, on reconstruction quality measured using {\em accuracy\/}, as explained in Section~\ref{sec:eva_integration}.}}
    \centering
    \label{table:comparison_recon_acc}
    \begin{tabular}{cccccccccccc}
        \toprule
        \multirow{2}{*}{Proxy} & \multirow{2}{*}{Recon.} & \multicolumn{5}{c}{Zhou et al. Planner} & \multicolumn{5}{c}{Smith et al. Planner}                                                                                                                                                   \\ \cmidrule(r){3-7} \cmidrule(r){8-12}
                               &                         & Image (\#)                              & 70\%$\downarrow$                         & 80\%$\downarrow$ & 90\%$\downarrow$ & 95\%$\downarrow$ & Image (\#) & 70\%$\downarrow$ & 80\%$\downarrow$ & 90\%$\downarrow$ & 95\%$\downarrow$ \\ \hline\hline  %
        \multirow{3}{*}{Box}   & Smith                   & 1,770                                    & 0.0256                                   & 0.0391           & 0.0695           & 0.1085           & 806        & \rev{0.0280}           & \rev{0.0427}           & \rev{0.0769}           & \rev{0.1229}           \\               %
                               & Ours (w/o)              & 1,717                                    & 0.0246                                   & 0.0378           & 0.0675           & 0.1038           & 699        & \rev{0.0253}           & \rev{0.0384}           & \rev{0.0680}           & \rev{0.1052}           \\               %
                               & Ours                    & 1,610                                    & \B{0.0186}                               & \B{0.0289}       & \B{0.0531}       & \B{0.0844}       & 796        & \rev{\B{0.0218} }      & \rev{\B{0.0346}}       & \rev{\B{0.0650}}       & \rev{\B{0.1044}}       \\ \hline        %
        \multirow{3}{*}{Inter} & Smith                   & 714                                     & 0.0193                                   & 0.0298           & 0.0545           & 0.0860           & 825        & \rev{0.0236}           & \rev{0.0368}           & \rev{0.0686}           & \rev{0.1110}           \\               %
                               & Ours (w/o)              & 747                                     & \B{0.0187}                               & \B{0.0291}       & \B{0.0536}       & \B{0.0848}       & 747        & \rev{0.0227}           & \rev{0.0351}           & \rev{0.0627}           & \rev{0.0987}           \\               %
                               & Ours                    & 696                                     & 0.0189                                   & 0.0297           & 0.0543           & 0.0850           & 696        & \rev{\B{0.0185}}       & \rev{\B{0.0303}}       & \rev{\B{0.0573}}       & \rev{\B{0.0887}}       \\               %
        \bottomrule
    \end{tabular}
\end{table*}

\begin{table*}[t!]
    \newcommand{\tabincell}[2]{\begin{tabular}{@{}#1@{}}#2\end{tabular}}
    \newcommand{\mcl}[1]{\multicolumn{1}{|c|}{#1}}
    \newcommand{\tg}[1]{\multirow{6}{*}{\rotatebox{90}{#1}}}
    \newcommand{\B}[1]{\textbf{#1}}
    \caption{\rev{Quantitative comparison between the different planners, on reconstruction quality measured using {\em completeness\/}, as explained in Section~\ref{sec:eva_integration}.}}
    \centering
    \label{table:comparison_recon_com}
    \begin{tabular}{cccccccccccc}
        \toprule
        \multirow{2}{*}{Proxy} & \multirow{2}{*}{Recon.} & \multicolumn{5}{c}{Zhou et al. Planner} & \multicolumn{5}{c}{Smith et al. Planner}                                                                                                                                                   \\ \cmidrule(r){3-7} \cmidrule(r){8-12}
                               &                         & Image (\#)                              & 70\%$\downarrow$                         & 80\%$\downarrow$ & 90\%$\downarrow$ & 95\%$\downarrow$ & Image (\#) & 70\%$\downarrow$ & 80\%$\downarrow$ & 90\%$\downarrow$ & 95\%$\downarrow$ \\ \hline\hline   %
        \multirow{3}{*}{Box}   & Smith                   & 1,770                                    & 0.2409                                   & 0.4607           & 0.9519           & 2.5048           & 806        & 0.3867           & 0.6710           & \B{1.1889}       & \B{2.6265}       \\                %
                               & Ours (w/o)              & 1,717                                    & \B{0.2352}                               & \B{0.4203}       & \B{0.8960}       & \B{1.8019}       & 699        & \B{0.3265}       & \B{0.6027}       & 1.1894           & 2.6428           \\                %
                               & Ours                    & 1,610                                    & 0.2576                                   & 0.4968           & 1.0011           & 2.4704           & 796        & 0.3928           & 0.6794           & 1.1870           & 2.6571           \\ \hline         %
        \multirow{3}{*}{Inter} & Smith                   & 714                                     & \B{0.2471}                               & \B{0.4789}       & \B{0.9611}       & 2.3776           & 825        & 0.3083           & 0.5661           & 1.1051           & 2.6055           \\                %
                               & Ours (w/o)              & 747                                     & 0.2549                                   & 0.4929           & 0.9898           & 2.4303           & 747        & \B{0.2856}       & \B{0.5356}       & 1.1274           & 2.6525           \\                %
                               & Ours                    & 696                                     & 0.2594                                   & 0.5013           & 0.9917           & \B{2.3767}       & 696        & 0.2969           & 0.5475           & \B{1.0382}       & \B{2.5033}       \\                %
        \bottomrule
    \end{tabular}
\end{table*}
\begin{table*}[t!]
    \newcommand{\tabincell}[2]{\begin{tabular}{@{}#1@{}}#2\end{tabular}}
    \newcommand{\mcl}[1]{\multicolumn{1}{|c|}{#1}}
    \newcommand{\tg}[1]{\multirow{6}{*}{\rotatebox{90}{#1}}}
    \newcommand{\B}[1]{\textbf{#1}}
    \caption{Comparing our new view planner with those from~\cite{Wolfgang18SIGA,DroneScan20} using F-score, Precision, and Recall. All planners employ our reconstruction predictor with the help of image features.}
    \centering
    \label{table:comparison_planner_Fscore}
    \begin{tabular}{ccccccccc}
        \toprule
        \multirow{2}{*}{Planner} & \multicolumn{4}{c}{Box Proxy} & \multicolumn{4}{c}{Inter Proxy}                                                                                                               \\ \cmidrule(r){2-5} \cmidrule(r){6-9}
                                 & Image (\#)                         & F-score$\uparrow$               & Precision$\uparrow$ & Recall$\uparrow$ &   Image (\#) & F-score$\uparrow$ & Precision$\uparrow$ & Recall$\uparrow$ \\ \hline\hline
        Smith et al.             & 735                           & 28.1596                         & 48.5530             & 19.8303          & 714   & 32.4972           & 53.8454             & 23.2709          \\
        Zhou et al.              & 1610                          & 34.5483                         & 52.4389             & \B{25.7598}      & 696   & 34.4410           & 52.1732             & \B{25.7047}      \\
        Ours                     & 764                           & \B{34.9627}                     & \B{58.0929}         & 25.0062          & 754   & \B{35.0951}       & \B{59.8070}         & 24.8339          \\ \hline
        \bottomrule
    \end{tabular}
\end{table*}

\begin{table*}[t!]
    \newcommand{\tabincell}[2]{\begin{tabular}{@{}#1@{}}#2\end{tabular}}
    \newcommand{\mcl}[1]{\multicolumn{1}{|c|}{#1}}
    \newcommand{\tg}[1]{\multirow{6}{*}{\rotatebox{90}{#1}}}
    \newcommand{\B}[1]{\textbf{#1}}
    \caption{Comparing our new view planner with those from~\cite{Wolfgang18SIGA,DroneScan20} using {\em accuracy\/}. All planners employ our reconstruction predictor and image features.}
    \centering
    \label{table:comparison_planner_Acc}
    \begin{tabular}{ccccccccccc}
        \toprule
        \multirow{2}{*}{Planner} & \multicolumn{5}{c}{Box Proxy} & \multicolumn{5}{c}{Inter Proxy}                                                                                                                                              \\ \cmidrule(r){2-6} \cmidrule(r){7-11}
                                 & Image (\#)                         & 70\%$\downarrow$                & 80\%$\downarrow$ & 90\%$\downarrow$ & 95\%$\downarrow$ &  Image (\#) & 70\%$\downarrow$ & 80\%$\downarrow$ & 90\%$\downarrow$ & 95\%$\downarrow$ \\ \hline\hline   %
        Smith et al.             & 735                           & 0.0218                          & 0.0346           & 0.0649           & 0.1044           & 714   & 0.0185           & 0.0303           & 0.0573           & 0.0888           \\                %
        Zhou et al.              & 1610                          & 0.0186                          & 0.0289           & 0.0531           & 0.0844           & 696   & 0.0189           & 0.0297           & 0.0544           & 0.0850           \\                %
        Ours                     & 764                           & \B{0.0153}                      & \B{0.0240}       & \B{0.0444}       & \B{0.0721}       & 754   & \B{0.0147}       & \B{0.0235}       & \B{0.0448}       & \B{0.0740}       \\                %
        \bottomrule
    \end{tabular}
\end{table*}

\begin{table*}[t!]
    \newcommand{\tabincell}[2]{\begin{tabular}{@{}#1@{}}#2\end{tabular}}
    \newcommand{\mcl}[1]{\multicolumn{1}{|c|}{#1}}
    \newcommand{\tg}[1]{\multirow{6}{*}{\rotatebox{90}{#1}}}
    \newcommand{\B}[1]{\textbf{#1}}
    \caption{Comparing our new view planner with those from~\cite{Wolfgang18SIGA,DroneScan20} using {\em completeness\/}. All planners employ our reconstruction predictor and image features.}
    \centering
    \label{table:comparison_planner_Comp}
    \begin{tabular}{ccccccccccc}
        \toprule
        \multirow{2}{*}{Planner} & \multicolumn{5}{c}{Box Proxy} & \multicolumn{5}{c}{Inter Proxy}                                                                                                                                              \\ \cmidrule(r){2-6} \cmidrule(r){7-11}
                                 & Image (\#)                        & 70\%$\downarrow$                & 80\%$\downarrow$ & 90\%$\downarrow$ & 95\%$\downarrow$ &  Image (\#) & 70\%$\downarrow$ & 80\%$\downarrow$ & 90\%$\downarrow$ & 95\%$\downarrow$ \\ \hline\hline  %
        Smith et al.             & 735                           & 0.3928                          & 0.6794           & 1.1870           & 2.6571           & 714   & 0.2969           & 0.5475           & 1.0382           & 2.5033           \\               %
        Zhou et al.              & 1610                          & \B{0.2576}                      & \B{0.4968}       & \B{1.0011}       & \B{2.4704}       & 696   & \B{0.2594}       & \B{0.5013}       & \B{0.9917}       & \B{2.3767}       \\               %
        Ours                     & 764                           & 0.3139                          & 0.5971           & 1.1862           & 2.6151           & 754   & 0.3188           & 0.5959           & 1.1362           & 2.4887           \\               %
        \bottomrule
    \end{tabular}
\end{table*}

\begin{table*}[t!]
    \newcommand{\tabincell}[2]{\begin{tabular}{@{}#1@{}}#2\end{tabular}}
    \newcommand{\mcl}[1]{\multicolumn{1}{|c|}{#1}}
    \newcommand{\tg}[1]{\multirow{4}{*}{\rotatebox{90}{#1}}}
    \newcommand{\B}[1]{\textbf{#1}}
    \caption{\rev{Quantitative comparisons on reconstruction quality, in terms of {\em accuracy\/} and {\em completeness\/}, between different methods using another reconstruction benchmark in~\cite{Wolfgang18SIGA}. Note that both test scenes are unseen by our data-driven reconstructability predictor. Our view planner is clearly the best performing one. In the table, we highlight the best performing numbers in each column in bold, and the second best performing numbers in italic.}}
    \centering
    \label{table:ny_uk}
    \begin{tabular}{ccccccccccccc}
        \toprule
        \multirow{2}{*}{Method} &                 & Acc.$\downarrow$ & Acc.$\downarrow$ & Comp.$\uparrow$ & Comp.$\uparrow$ & Comp.$\uparrow$ &                 & Acc.$\downarrow$ & Acc.$\downarrow$ & Comp.$\uparrow$ & Comp.$\uparrow$ & Comp.$\uparrow$ \\
                                &                 & 90\%             & 95\%             & 0.02m           & 0.05m           & 0.075m          &                 & 90\%             & 95\%             & 0.02m           & 0.05m           & 0.075m          \\ \hline\hline
        Smith et al.            & \mcl{\tg{NY-1}} & 0.053            & 0.792            & 36.010          & 44.740          & 49.470          & \mcl{\tg{UK-1}} & \B{0.028}        & {\em 0.051\/}            & \B{32.040}      & {\em 37.740\/}          & {\em 40.620\/}          \\
        Zhou et al.             & \mcl{}          & {\em 0.030\/}            & 0.342            & \B{38.190}      & {\em 45.220\/}          & {\em 49.780\/}          & \mcl{}          & {\em 0.030\/}            & 0.054            & 30.750          & 35.960          & 38.770          \\
        Liu et al.              & \mcl{}          & \B{0.029}        & \B{0.107}        & N/A               & 44.72           & 49.33           & \mcl{}          & \B{0.028}        & 0.052            & N/A               & 36.71           & 39.58           \\
        Ours                    & \mcl{}          & 0.039            & {\em 0.192\/}            & {\em 38.077\/}          & \B{46.046}      & \B{50.523}      & \mcl{}          & 0.031            & \B{0.050}        & {\em 31.755\/}          & \B{37.843}      & \B{40.795}      \\
        \bottomrule
    \end{tabular}
\end{table*}
\begin{table*}[t!]
    \newcommand{\B}[1]{\textbf{#1}}
    \caption{Comparing our new path planer with state-of-the-art alternatives~\cite{Wolfgang18SIGA,DroneScan20} on two real scenes. Each real scene has a high precision LiDAR capture as the ground truth model. We report F-score, accuracy, and completeness of the final reconstruction results.}
    \centering
    \label{table:realscene}
    \begin{tabular}{cccccccccccc}
        \toprule
        \multirow{2}{*}{Scene}    & \multirow{2}{*}{Method} & \multirow{2}{*}{Image (\#)} & \multirow{2}{*}{F-score$\uparrow$} & \multirow{2}{*}{Precision$\uparrow$} & \multirow{2}{*}{Recall$\uparrow$} & Acc.$\downarrow$ & Acc.$\downarrow$ & Acc.$\downarrow$ & Comp.$\downarrow$ & Comp.$\downarrow$ & Comp.$\downarrow$ \\
                                  &                         &                        &                                    &                                      &                                   & 80\%             & 90\%             & 95\%             & 80\%              & 90\%              & 95\%              \\ \hline\hline
        \multirow{3}{*}{Polytech} & Smith et al.            & 678                    & 9.9549                             & 7.1530                               & 16.3659                           & 0.2106           & 0.4385           & 0.7271           & 0.3011            & 0.8440            & 2.0124            \\
                                  & Zhou et al.             & 1364                   & 10.3966                            & 7.3969                               & 17.4888                           & 0.1872           & 0.4229           & 0.7108           & 0.2775            & 0.8348            & 1.8146            \\
                                  & Ours                    & 1141                   & \B{16.5603}                        & \B{11.8909}                          & \B{27.2685}                       & \B{0.1435}       & \B{0.3331}       & \B{0.6368}       & \B{0.2368}        & \B{0.7448}        & \B{1.7062}        \\ \hline
        \multirow{3}{*}{ArtSci}   & Smith et al.            & 2900                   & 8.4057                             & 5.7568                               & 15.5701                           & 0.2338           & 0.4066           & 0.6450           & 0.5787            & 1.5299            & 2.5265            \\
                                  & Zhou et al.             & 3286                   & 11.5342                            & 8.1013                               & 20.0158                           & \B{0.1965}       & \B{0.3517}       & \B{0.5494}       & 0.5650            & 1.5680            & 2.5586            \\
                                  & Ours                    & 2600                   & \B{13.7585}                        & \B{9.1734}                           & \B{27.5069}                       & 0.2167           & 0.3984           & 0.6440           & \B{0.4632}        & \B{1.3959}        & \B{2.3358}        \\
        \bottomrule
    \end{tabular}
\end{table*}

\begin{figure*}[t!]
	\centering
	\includegraphics[width=1\linewidth]{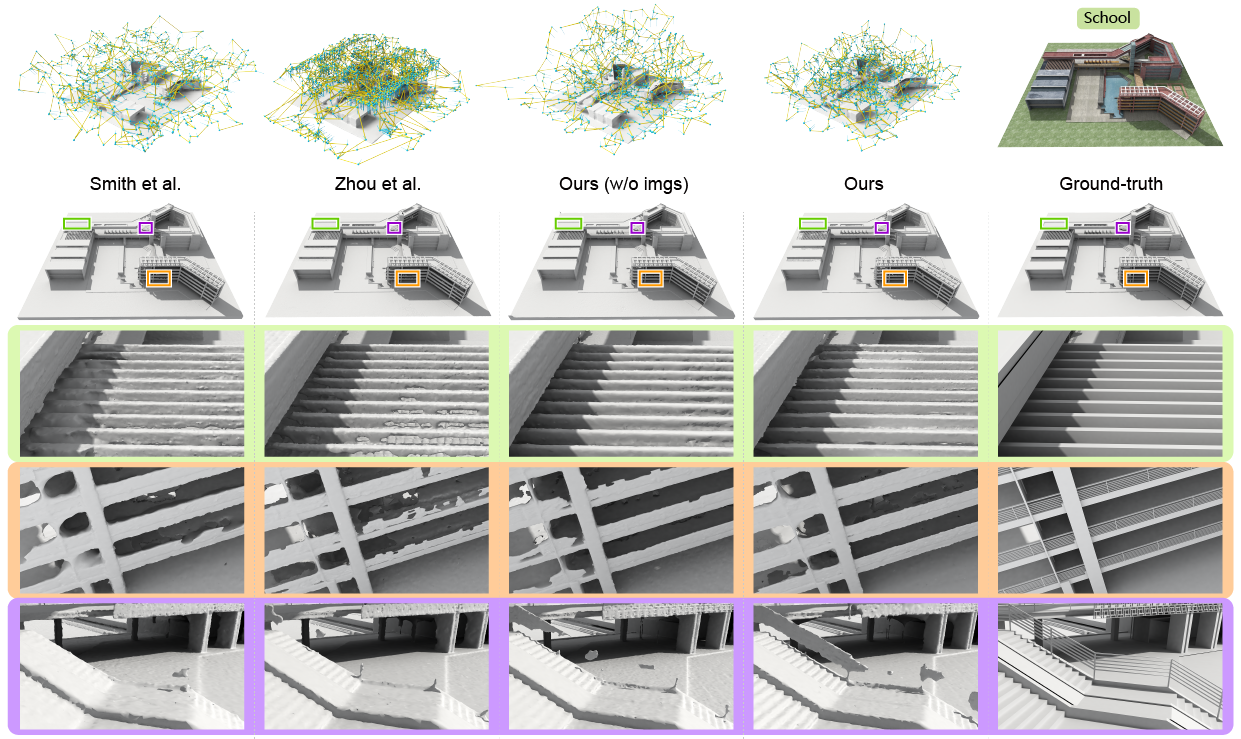}
	\caption{A visual comparison of reconstruction results produced by different methods: Smith et al.~\shortcite{Wolfgang18SIGA}, Zhou et al.~\shortcite{DroneScan20}, 
		and ours (with vs.~without using image features). With the ground truth meshes as references, we can see that the reconstruction results obtained by our method can more faithfully recover geometric details and sharp features.}
	\label{fig:school_recon}
\end{figure*}

\subsection{Reconstructability}
\label{sec:eva_recon}

We compare our reconstructability predictor with the heuristic estimate from Smith et al.~\shortcite{Wolfgang18SIGA} using
the \textit{Spearman's rank correlation coefficient}, which is a measurement to quantify correlations. Indeed, a quality 
reconstructability estimate or prediction should output values that best reflect the final reconstruction quality.
Specifically, the highest {\em Spearman correlation factor} is obtained when the order of the predicted reconstructability is the same 
as the order of the reconstruction quality in a scene. 

Our experiments have been conducted using the trajectories and corresponding reconstruction results from the UrbanScene3D dataset~\cite{UrbanScene3D}. Specifically, we train our network on three scenes: \textit{Castle, Village, Bridge}, as shown in Fig.~\ref{fig:eva_training_set}, with testing done on a new scene, \textit{School}. We report the Spearman correlation of \textit{Inter, Fine} proxy for phase 1 and all four levels of proxy \textit{Box, Coarse, Inter, Fine} for phase 2. The whole test set contains 24 trajectories and their corresponding reconstruction results. These trajectories share different characteristics, e.g., in terms of flight patterns and view density. We use the reconstructability calculated by Smith et al.~\shortcite{Wolfgang18SIGA} and the number of visible views as baselines for comparison.

\paragraph{Reconstructability predicted by Smith et al.~\shortcite{Wolfgang18SIGA}} 
For each sample point, we extract the visible viewpoints and calculate the corresponding reconstructability. We use the default parameter $k1=32$, $k3=8$, $alpha1=\pi / 16$, $alpha3=\pi / 4$, as in their paper. 

\paragraph{Number of visible views} 
Alternatively, with the intuition that sample points with more visible viewpoints tend to lead to higher reconstruction quality, we also calculate the number of visible viewpoints for each sample point and compute its Spearman correlation factor with respect to the final reconstruction quality.

As shown in Table~\ref{table:spearman_p0}, the values predicted by our reconstructability predictor have a higher Spearman correlation factor than those obtained by the other baselines. Furthermore, the margin of gains by our method is even greater when the proxy is inaccurate, as can be seen from Table~\ref{table:spearman_p1}. Note that the \textit{Inter} level of proxy refers to the proxy generated by the rapid pre-flight using \textit{Oblique Photography}. \textit{Coarse} and \textit{Box} proxies refer to those extracted by satellite images. Both of these image acquisition and scene reconstruction workflows have been common practices in the industry.

In Fig.~\ref{fig:eva_contribution}, we show the weight of each viewpoint when predicting reconstructability. While the contribution of each viewpoint is automatically extracted by our network without any supervision, there are still observable patterns, e.g., larger viewing distance tends to result in lower contribution; viewpoints with an appropriate baseline and scale difference often produce higher contributions. 

\begin{figure*}[t!]
	\centering
	\includegraphics[width=0.95\linewidth]{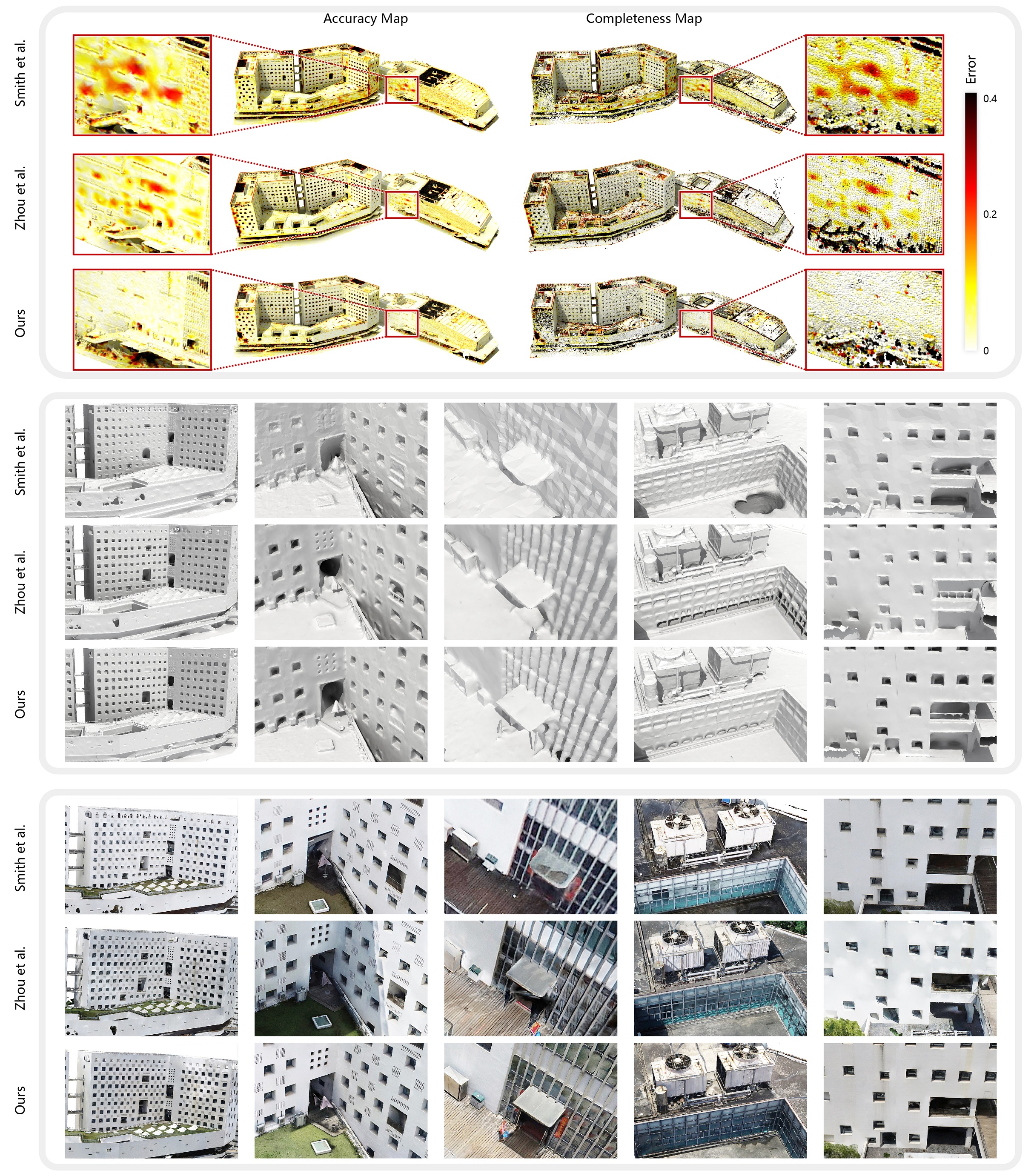}
	\caption{\rev{Qualitative results on the \textit{Polytech} scene, compared with previous planning and reconstruction methods~\cite{Wolfgang18SIGA,DroneScan20}. \textit{Top:} Global and Local error map on the scene. The accuracy map is collected by calculating the shortest distance from the reconstructed mesh to the GT LiDAR points, while the completeness map is collected by collecting the shortest distance from the ground truth LiDAR points to the reconstructed mesh. \textit{Middle:} Local geometry details on the reconstructed mesh. \textit{Bottom:} Local textured details on the reconstructed mesh.}}
	\label{fig:l7_detail}
\end{figure*}

\begin{figure*}[t!]
	\centering
	\includegraphics[width=0.95\linewidth]{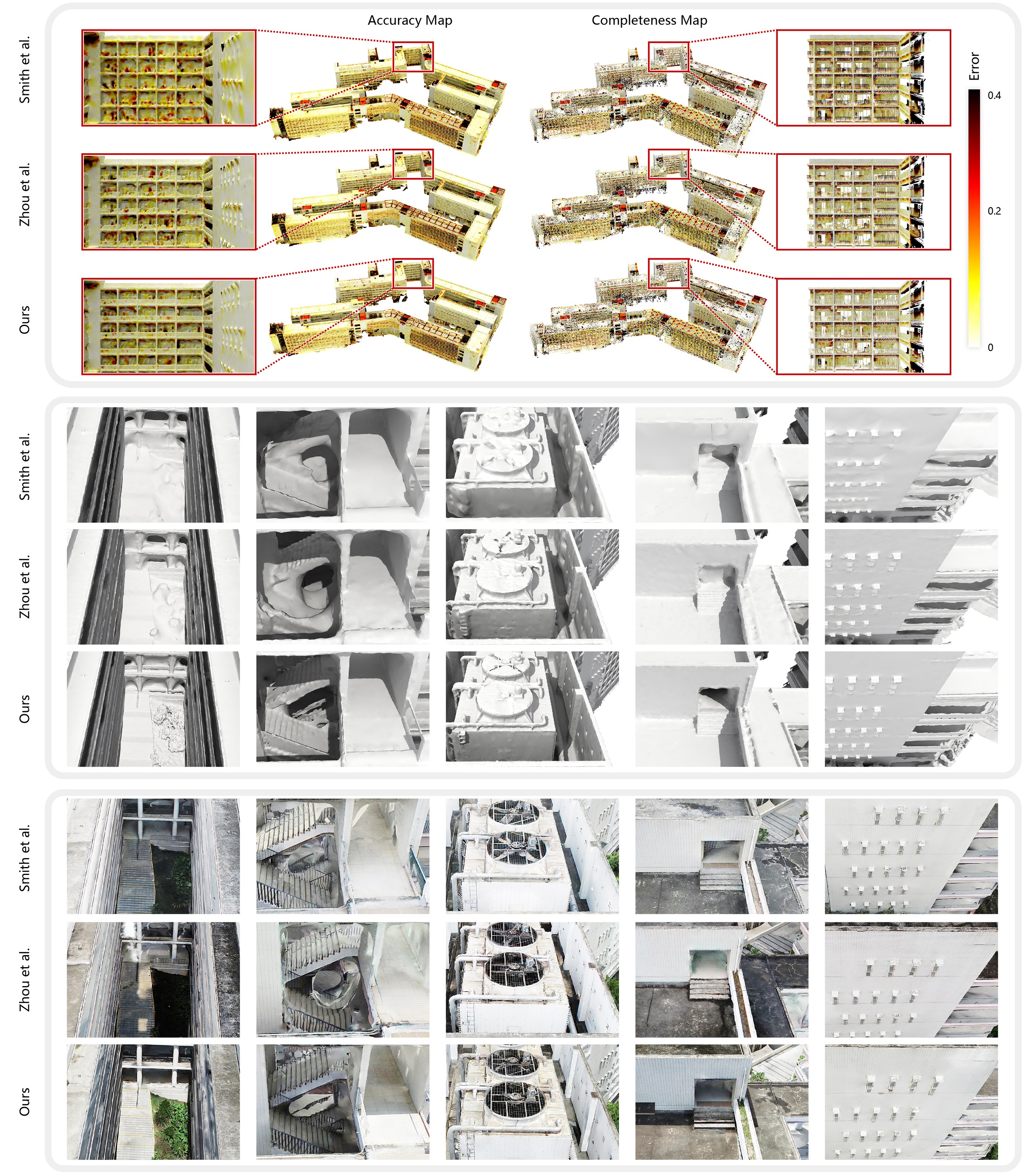}
	\caption{\rev{Visualization of the reconstruction results on the \textit{ArtSci} scene compared with previous methods~\cite{Wolfgang18SIGA,DroneScan20}. Unlike \textit{Polytech}, \textit{ArtSci} contains two irregular buildings with more complex geometries, leading to increased difficulty towards trajectory planning. We also show the LiDAR points that were collected by the high-resolution LiDAR scanner, which are used as the ground truth model for evaluation.}}
	\label{fig:hwl_detail}
\end{figure*}

\begin{figure*}[t!]
	\centering
	\includegraphics[width=0.92\linewidth]{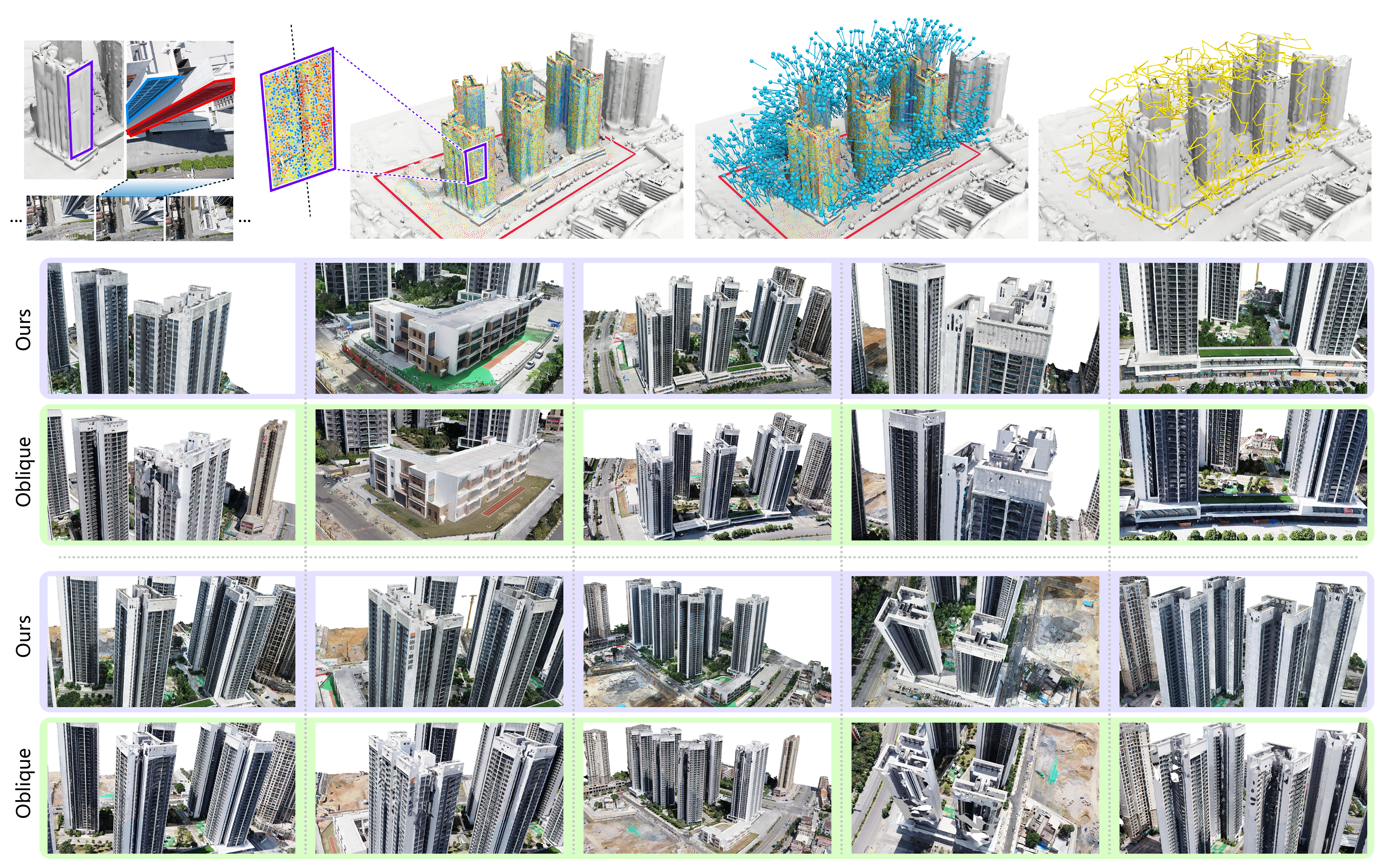}
	\caption{Qualitative results on the \textit{Apartment} scene, compared with \textit{Oblique Photography}, a widely-used path planning method in the industry. \textit{Apartment} contains six high-rise buildings (higher than 80m), making it difficult to obtain a quality reconstruction using \textit{Oblique Photography}. Moreover, our method can extract and incorporate scene uncertainties from input images into reconstructability prediction. As shown in the top left of the image, sample points located over inaccurate regions of the proxy can be implicitly identified by our reconstructability predictor. They have lower reconstructability, which would encourage the view planner to produce viewpoints with large baselines in order to obtain a comprehensive observation over these inaccurate regions.}
	\label{fig:ronghu}
\end{figure*}

\subsection{Integration with Existing Planners}
\label{sec:eva_integration}

We simulate a complete scene reconstruction pipeline to evaluate the integration of our reconstructability predictor with existing path planners~\cite{Wolfgang18SIGA,DroneScan20}, after slight modifications as discussed in Sec.~\ref{sec:planner}. These newly adapted
planners are compared to their counterparts with the reconstructability estimated by the approach in Smith et al.~\shortcite{Wolfgang18SIGA}.
Similar to Sec.~\ref{sec:eva_recon}, we test reconstructability prediction and path planning on the \textit{School} scene. Also, we use two proxy levels, \textit{Coarse} and \textit{Inter}, to plan trajectories as they represent two common ways to obtain scene proxies in practice: a quick reconstruction from a rapid flight and 2.5D box extraction from satellite images, respectively.

\rev{For evaluation, two different metrics, accuracy and completeness, are employed to compare quality of the reconstructed meshes, as in Smith et al.~\shortcite{Wolfgang18SIGA}. While accuracy measures how close a reconstructed mesh is to the ground truth mesh, completeness reveals how well the ground truth mesh is ``covered'' by the reconstruction. In other words, accuracy accounts from distances from the reconstructed mesh to the ground truth mesh, while completeness is measured based on distances from the ground truth mesh to the reconstructed mesh.} \rev{Specifically, for each sample point on the surface of a reconstructed mesh, we find the closest point on the ground truth mesh and compute their distance. We then sort these distances for all the sample points and report an accuracy number $y$, with respect to a given percentage $x\%$, if exactly $x\%$ of the distances are less than $y$. Clearly, a lower accuracy number would corroborate with higher reconstruction quality. For completeness, we simply switch the roles of the reconstructed mesh and the ground truth mesh and perform the same computations.} Additionally, we report F-scores~\cite{Knapitsch2017} as a summary metric, where a threshold of $10cm$ was applied to extract inliers and outliers.

\rev{Quantitative comparisons on reconstruction quality between the different planners are shown in Tables~\ref{table:comparison_recon_fpr}, \ref{table:comparison_recon_acc}, and \ref{table:comparison_recon_com}. Specifically, Table~\ref{table:comparison_recon_fpr} shows that, in terms of F-score, Precision, and Recall, the planners employing our reconstructability predictor, either with image inputs or without, generally outperform the baseline planners guided by reconstructability estimates from Smith et al.~\shortcite{Wolfgang18SIGA}. Comparisons in terms of accuracy and completeness measures also exhibit a similar trend, as shown in Tables~\ref{table:comparison_recon_acc} and \ref{table:comparison_recon_com}.}

\subsection{Integration with the New Planner}
\label{sec:eva_planner}

\rev{To evaluate the performance of our new view planner, we compare it to the planners from~\cite{Wolfgang18SIGA,DroneScan20}, all using our reconstructability predictor 
with the help of image features. We follow the same evaluation strategy as in Sec.~\ref{sec:eva_integration}, also testing on the \textit{School} scene. The quantitative
results in Tables~\ref{table:comparison_planner_Fscore}, \ref{table:comparison_planner_Acc}, and~\ref{table:comparison_planner_Comp} show that our planner 
generally outperforms the alternatives, attaining higher precision while maintaining a comparable recall on the final reconstruction. Fig.~\ref{fig:school_recon} provides 
a visual comparison between reconstruction results produced by the different methods.}

We also evaluate our method using the scene reconstruction benchmark proposed by Smith et al.~\shortcite{Wolfgang18SIGA}. Following Zhou et al.~\shortcite{DroneScan20}, we choose \textit{NY-1} and \textit{UK-1} as the test scenes, since \textit{Bridge-1} has been used for training. Note that \rev{both of these test scenes are new to our reconstructability predictor.} As the results in Table~\ref{table:ny_uk} show, \rev{while our new planner does not come on top in every reported measure,
it is clearly the best performing one among the four methods compared. Note that to strictly follow the measures employed by the reconstruction benchmark, the reported
completeness measure in Table~\ref{table:ny_uk} is different from those shown in the other tables: the roles of $x$ and $y$ in the prior completeness definition are switched so that larger numbers reflect higher reconstruction quality.}

\subsection{Test on Real Scenes}
\label{sec:eva_real}

To further demonstrate the performance of our method, we show qualitative reconstruction results obtained by our new view planner on several challenging real 3D scenes. We chose three scenes possessing different scales and characteristics. Specifically, \textit{Polytech} contains two buildings with weak texture, repeated patterns, and large height differences, while \textit{ArtSci} contains two irregular buildings with complex geometries and occlusions. Finally, \textit{Apartment} contains six high-rise buildings covering the largest area.

We compare our reconstruction results for \textit{Polytech} and \textit{ArtSci} with the high resolution LiDAR point clouds from UrbanScene3D~\cite{UrbanScene3D} as ground truth. As shown in Table~\ref{table:realscene}, the scenes reconstructed by our method are generally of superior quality. Note that our model was trained using synthetic data, without any fine-tuning on real scenes. Fig.~\ref{fig:l7_detail} and Fig.~\ref{fig:hwl_detail} show visual comparisons, demonstrating that our reconstruction results are more accurate, especially over regions with complex geometries and occlusions.

Since we do not have a ground-truth mesh for \textit{Apartment}, we simply compare our reconstruction results qualitatively with those obtained by a widely-adopted industrial algorithm: \textit{Oblique Photography}. As demonstrated in Fig.~\ref{fig:ronghu}, our method achieves better reconstruction quality visually, especially over regions near the ground, where images from \textit{Oblique Photography} can hardly be observed. Moreover, we show that our reconstructability predictor can extract the potential inaccurate geometry from the images and propagate the uncertainty to the reconstructability predictor.

\section{Conclusion, Limitation, and Future Work} 
\label{sec:future}

Our work shows that reconstructability, in the context of drone path planing for urban scene acquisition, is a learnable measure. Specifically, we
define reconstructability, i.e., the expected scene reconstruction quality, as a function of proxy geometry and a set of viewpoints, and introduce the
first data-driven predictor trained on synthetic data from the new UrbanScene3D dataset. 

\rev{While our acquisition problem falls under the general realm of ``learning to reconstruct'', it differs from most reconstructive tasks including 
all recent works on neural fields~\cite{NeuralFields}. In these latter works, and under the typical setting for 3D reconstruction, 
the input at test time is a direct, albeit under-, representation of the target 3D scene, e.g., a shape code for IM-Net~\cite{IMNet} or multi-view images for NeRF~\cite{NeRF}. 
In our problem setting, such inputs are not given; we must predict a view plan to acquire these inputs first, on-the-fly during test time, and then reconstruct the scene.
As a result, the design of our learning framework has to handle at least two gaps: the domain gap between synthetic and real data, and the accuracy gap between the proxy and true geometry of the reconstructed scene.}

Extensive experiments have been conducted to demonstrate that our learned reconstructability better correlates with the true reconstruction quality 
than existing heuristic estimates. Combined with an iterative view optimization scheme, our predictor can be integrated into both previous and our 
new path planners, leading to consistent improvements on reconstruction quality. Qualitative and quantitative results are presented for both 
synthetic and real scenes to demonstrate generalizability of our learning framework.

As a first attempt at a learning framework for reconstructability, our work still has several limitations. For example, while our feature learning
is geometry- and uncertainty-aware, it does not explicitly account for material properties of the acquired scene. More critically, since 
our predictor operates on point-view and point-image features that are both defined on the scene proxy, the scale and variability of the 
gaps between the proxy and the true scene geometry can impact the quality of the learned model. Currently, we relate points from the
proxy and the true surface via closest distances, which is a simple heuristic but not a reliable correspondence.

Furthermore, both the view eliminator from Zhou et al.~\shortcite{DroneScan20} and the view adjuster from Smith et al.~\shortcite{Wolfgang18SIGA} 
need a scoring function to transform the calculated reconstructability from the sample points to the viewpoint, as the view planner must decide 
whether a viewpoint is redundant or a new choice is better. Our current scoring function is quite heuristic, e.g., Zhou et al.~\shortcite{DroneScan20} 
use the smallest reconstructability of the sample point that the viewpoint can observe as the score. Compared with the existing reconstructability 
definitions, one may also consider how to directly define the reconstructability on the optimized viewpoints.

Besides addressing the above limitations, it would also be interesting to explore the correlation between viewpoints in the MVS pipeline. Learning 
such correlations is useful for view planning, pose estimation, as well as sparse/dense urban reconstruction. Finally, we
are interested in applying and adapting our learning framework for robot-assisted 3D object or indoor scene acquisition.

\section*{Acknowledgments}
We thank the anonymous reviewers for their valuable feedback.
This work was supported in part by NSFC (U2001206, 62161146005, U21B2023), Guangdong Talent Program (2019JC05X328), RGC HKSAR (CUHK14206320), NSERC (611370),  Shenzhen Science and Technology Program (KQTD20210811090044003, RCJC20200714114435012, JCYJ20210324120213036), and Guangdong Laboratory of Artificial Intelligence and Digital Economy (SZ).

\bibliographystyle{ACM-Reference-Format}
\bibliography{DroneRecon}

\end{document}